\documentclass[twocolumn]{aastex6}

\usepackage{graphicx}
\usepackage{array}
\usepackage{amssymb}
\usepackage{natbib}
\usepackage{url}
\usepackage[utf8]{inputenc}  
\usepackage{amsmath}  
\usepackage{comment}
\usepackage{footnote}

\newcommand{\degree}{\ensuremath{^\circ}}
\renewcommand*{\thefootnote}{\fnsymbol{footnote}}

\shorttitle{Outflows in Type 1 AGNs}
\shortauthors{Rakshit and Woo}

\begin{document}


\title{A census of ionized gas outflows in type 1 AGNs: gas outflows in AGNs. V}



\author{Suvendu Rakshit and Jong-Hak Woo}
\affil{Astronomy Program, Department of Physics and Astronomy, Seoul National University, Seoul 151-742, Republic of Korea}

\email{woo@astro.snu.ac.kr}

\begin{abstract}
We present a systematic study of ionized gas outflows based on the velocity shift and dispersion of the [O III] $\lambda 5007$\AA \, emission line, using a sample of $\sim 5000$ Type 1 AGNs at $z<0.3$ selected from Sloan Digital Sky Survey. This analysis is supplemented by the gas kinematics of Type 2 AGNs from \citet{2016ApJ...817..108W}. For the majority of Type 1 AGNs (i.e., $\sim 89$\%), the [O III] line profile is best represented by a double Gaussian model, presenting the kinematic signature of the non-virial motion.
Blueshifted [O III] is more frequently detected than redshifted [O III] by a factor of 3.6 in Type 1 AGNs, while the ratio between blueshifted to redshifted [O III] is only 1.08 in Type 2 AGNs due to the projection and orientation effect. 
The fraction of AGNs with outflow signatures is found to increase steeply with [O III] luminosity and Eddington ratio, while Type 1 AGNs have larger velocity dispersion and more negative velocity shift than Type 2 AGNs. The [O III] velocity $-$ velocity dispersion (VVD) diagram of Type 1 AGNs expands towards higher values with increasing luminosity and Eddington ratio, suggesting that the radiation pressure or wind is the main driver of gas outflows, as similarly found in Type 2 AGNs. In contrast, the kinematics of gas outflows is not directly linked to the radio activity of AGN.
\end{abstract}





\section{Introduction}

Mass-accreting supermassive black holes are manifested as active galactic nuclei (AGNs), which are classified into two categories based on the unification model: Type 1 AGNs, of which the central engine is directly viewed, and Type 2 AGNs, of which the central engine is obscured by the torus \citep{1993ARA&A..31..473A,1995PASP..107..803U}.
The observed correlation between black hole mass and host galaxy properties is typically interpreted as that feeding and feedback work together in self-regulation of black hole growth and galaxy evolution \citep{2013ARA&A..51..511K}.  During the process of galaxy-galaxy interaction, for example, strong inflows supply a vast amount of gas to the supermassive black hole helping it to evolve and power AGN. The radiation emitted from AGN drives out a large amount of gas, quenching star formation and also the growth of the black hole \citep[see review by][]{2015ARA&A..53..115K}.

The feedback of AGNs is considered to be traced by frequently seen gas outflows in various scales \citep[see review by][]{2012ARA&A..50..455F}. Though various mechanisms, i.e., the disc wind \citep[e.g.,][]{2003ARA&A..41..117C,2009ApJ...702L.187R,2015Natur.519..436T}, radiation pressure on dust \citep{1998MNRAS.294L..47B,2002ApJ...572..753D,2010MNRAS.402.2211A}, interaction of radio jet with clouds \citep{2005MNRAS.359..781S,2008A&A...491..407N,2012ApJ...747...95G} etc. have been considered, the main driver of AGN feedback remains largely debatable \citep[see][]{2015ARA&A..53..115K}. AGN-driven gas outflows have been detected in various energy bands, e.g., optical, UV and X-ray \citep[e.g.,][]{2003ApJ...593L..65R,2013MNRAS.436.3286A} allowing to probe AGN feedback in different scales, while the kpc-scale outflows observed in the narrow line region (NLR) are particularly important in understanding AGN feedback since they are extended to galactic scales, where the outflows may interact with interstellar medium and suppress star formation.  

The [O III] $\lambda 5007$\AA \, emission line being strong in AGN spectra is a good tracer of outflows and consequently subjected to various studies. On one hand, spatially resolved spectroscopy based on [O III] kinematics mapped the velocity structure in the NLR \citep[e.g.,][]{2000ApJ...532L.101C,2013ApJS..209....1F}. More detailed studies of gas kinematics and star formation became possible thanks to the integral field spectroscopy \citep[e.g.,][]{2006ApJ...650..693N,2014MNRAS.441.3306H,2016ApJ...819..148K,2016ApJ...833..171K,2017ApJ...837...91B,2017MNRAS.467.2612W,2018ApJ...858...48M,2018MNRAS.476.2760F}. On the other hand, more systematic studies of [O III] kinematics were based on the spatially integrated spectra obtained from large surveys since a large sample of AGNs can be utilized, revealing that outflows are prevalent, particularly among luminous AGNs \citep[e.g.,][]{
2013MNRAS.433..622M,2014ApJ...795...30B,2016ApJ...817..108W,2016ApJ...831....7S,2017ApJ...839..120W,2017MNRAS.468..620Z,2018ApJ...852...26W,2018ApJ...856...76D}.

The kinematics of [O III] are mainly caused by AGN outflows, while the virial motion due to the gravitational potential of the host galaxy is partly responsible for the broadening of the [O III] line \citep{1984ApJ...281..525H,1996ApJ...465...96N,2008ApJ...680..926K,2014JKAS...47..167W,2017ApJ...839..120W,2017ApJ...842....5E}. 
A positive correlation between mid-infrared luminosity and velocity width of [O III] suggests that the gas outflows are mainly radiation driven  \citep{2014MNRAS.442..784Z}. 
The studies of gas outflows based on a large sample of Type 1 and Type 2 AGNs support this idea \citep[e.g.,][]{2016ApJ...817..108W, 2018ApJ...852...26W}.
\citet{2016ApJ...817..108W} studied outflow kinematics of $\sim 39,000$ Type 2 AGNs at $z<0.3$ by carefully estimating velocity dispersion and shift of [O III] with respect to the stellar velocity dispersion and systemic velocity of the host galaxies. Their combined analysis of velocity dispersion and velocity shift exhibits the presence of outflow signatures in the majority of the high luminous AGNs and the strong dependence of the outflow properties on the radiation emitted by AGNs.

In a series of papers on ionized gas outflows in AGNs, the demography of ionized gas outflows in type 2 AGNs was reported, respectively, based on [O III] \citep{2014ApJ...795...30B,2016ApJ...817..108W} and H$\alpha$ \citep{2017ApJ...845..131K}, while \citet{2016ApJ...828...97B} constrained the physical properties of the outflows, e.g., launching velocity, dust extinction, and the opening angle of the cone, based on the kinematical modeling of the outflows and Monte Carlo simulations. Based on these studies \citet{2017ApJ...839..120W} showed that AGNs with strong outflows tend to have a regular star formation rate similar to the star-forming galaxies in the main sequence, while AGNs with weak or no outflows have on average much lower specific star formation rate, suggesting that the effect of AGN-driven outflows is delayed. 

In this paper, we focus on the gas outflows of Type 1 AGNs based on the kinematics of [O III]. Compared to Type 2 AGNs, Type 1 AGNs have a number of merits. First, since the direction of the outflows is closer to the line-of-sight than that of Type 2 AGNs, the projection effect is less problematic and the measured velocity is expected to be higher. Second,  the main physical parameters of AGNs, i.e., black hole mass ($M_{\mathrm{BH}}$) and Eddington ratio ($\lambda_{\mathrm{Edd}}$) can be properly estimated in Type 1 AGNs, while for Type 2 AGNs, mass and bolometric luminosity is much more difficult to estimate due to the lack of broad emission lines and AGN continuum. 
On the other hand, the downside of type 1 AGNs includes the difficulty of measuring the systemic velocity (e.g., based on stellar 
absorption lines), which is required to measure the velocity shift of outflows based on gas emission lines, and host galaxy mass or stellar velocity dispersion, which are needed to calculate the host galaxy gravitational potential for removing the effect of the virial motion in the width of gas emission lines. Therefore, it is important to combine Type 1 and Type 2 AGNs for better understanding gas outflows and their connection to AGN energetics.

We investigated outflow kinematics based on [O III] for a large sample of Type 1 AGNs at $z<0.3$. By combining these Type 1 AGNs with the Type 2 AGNs from \citet{2016ApJ...817..108W}, we perform a demography of ionized gas outflows over a  luminosity range of $\sim 5$ orders of magnitude. The data and spectral analysis method are presented in Section \ref{sec:data}. The main results are given in Section \ref{sec:results} and discussion in Section \ref{sec:discussion}. The summary and conclusions are presented in Section \ref{sec:conclusion}. A cosmology with $H_0=70 \, \mathrm{km \, s^{-1} Mpc^{-1}}$, $\mathrm{\Omega_m}=0.3$ and $\Omega_{\Lambda}=0.7$ is used throughout the paper.


\section{ Data and Spectral analysis}\label{sec:data}
\subsection{Sample}
To select Type 1 AGNs, we used the ``specObj''\footnote{``specObj'' contains only the best spectra for any object obtained by SDSS}
 data products of the SDSS DR12 catalog \citep{2015ApJS..219...12A} and considered the objects that are classified as``QSO'' by the SDSS spectroscopic pipeline \citep{2002AJ....123.2945R}. We chose the objects having $z<0.3$ and a median signal-to-noise ratio (S/N) $>10$ pixel$^{-1}$ given by the SDSS pipeline. This initial sample includes 8645 sources, of which we carefully and systematically analyzed the spectra. 
 
\subsection{Spectral analysis}
Prior to the multicomponent spectral analysis, Galactic extinction was corrected for each spectrum, using the extinction map of \citet{1998ApJ...500..525S} and the Milky way extinction law with $R_V=3.1$ from \citet{1989ApJ...345..245C}. The spectra were then brought to the rest frame using the redshift provided by the SDSS pipeline.   

A multicomponent spectral fitting procedure was applied to model the observed spectra as used for our previous studies of Type 1 AGNs \citep{2015ApJ...799..164P,2015ApJ...801...38W,2018arXiv180402798W}. First, the continuum was modeled by a combination of a single power law ($f_{\lambda} = A \lambda^{\alpha_{\lambda}}$), a Fe II template and a host galaxy template, representing AGN continuum contribution, iron lines, and stellar contribution, respectively, in the continuum regions of $4430–4770$\AA \, and $5080–5450$\AA. During this step, all the narrow and broad emission lines were masked out. The Fe II template from \citet{2010ApJS..189...15K} was used since it is the best available template for accurately fitting various blended Fe II emission lines seen in AGN spectra \citep{2017ApJ...839...93P,2017ApJS..229...39R}. The stellar template was taken from the Indo-US spectral library\footnote{\url{https://www.noao.edu/cflib/}} \citep{2004ApJS..152..251V}, consisting of seven spectra of G and K type giant stars of various temperatures. To find the best-fit continuum model we performed nonlinear Levenberg-Marquardt least-squares minimization using {\sc IDL} fitting package {\sc mpfit}\footnote{\url{http://purl.com/net/mpfit}}\citep{2009ASPC..411..251M}. This allowed us to properly decompose all the components and estimate velocity shift and widths of the Gaussian broadening kernels used to convolve the host galaxy and Fe II templates. The host subtracted AGN continuum luminosity at $5100$\AA \, was then estimated using the AGN power law model. The best-fit model continuum was subtracted from the spectra prior to the modeling of the H$\beta$ emission line region.

The H$\beta$-[O III] line region consists of the H$\beta$ broad and narrow emission lines, [O III] $\lambda4687$, $5007$\AA \, doublets  and He II$\lambda4687$\AA. The H$\beta$ broad component was fitted using a sixth-order Gauss-Hermite series while the narrow component was fitted using a single Gaussian component. The upper limit in the full width at half maximum (FWHM) of the narrow H$\beta$ component was set to 1200 km s$^{-1}$. The HeII line was modeled using two Gaussian functions, while both $\lambda4687$, $5007$\AA \, doublets were modeled using two Gaussian functions; one for the core with an upper limit of 1200 km s$^{-1}$ and another for the wing. The flux ratios of [O III] doublets were fixed to their theoretical value. This multi-component fitting procedure was applied to all spectra, and then those having S/N at 5100\AA \, $>$10, and the amplitude-to-noise ratio (A/N) of H$\beta$ line $>5$ were chosen for further analysis. Note that these criteria were adopted to be consistent with the analysis of Type 2 AGNs by \citet{2016ApJ...817..108W}. As a result, we finalized the sample of 5717 Type 1 AGNs.  

\begin{figure}
\centering
\resizebox{9cm}{3.5cm}{\includegraphics{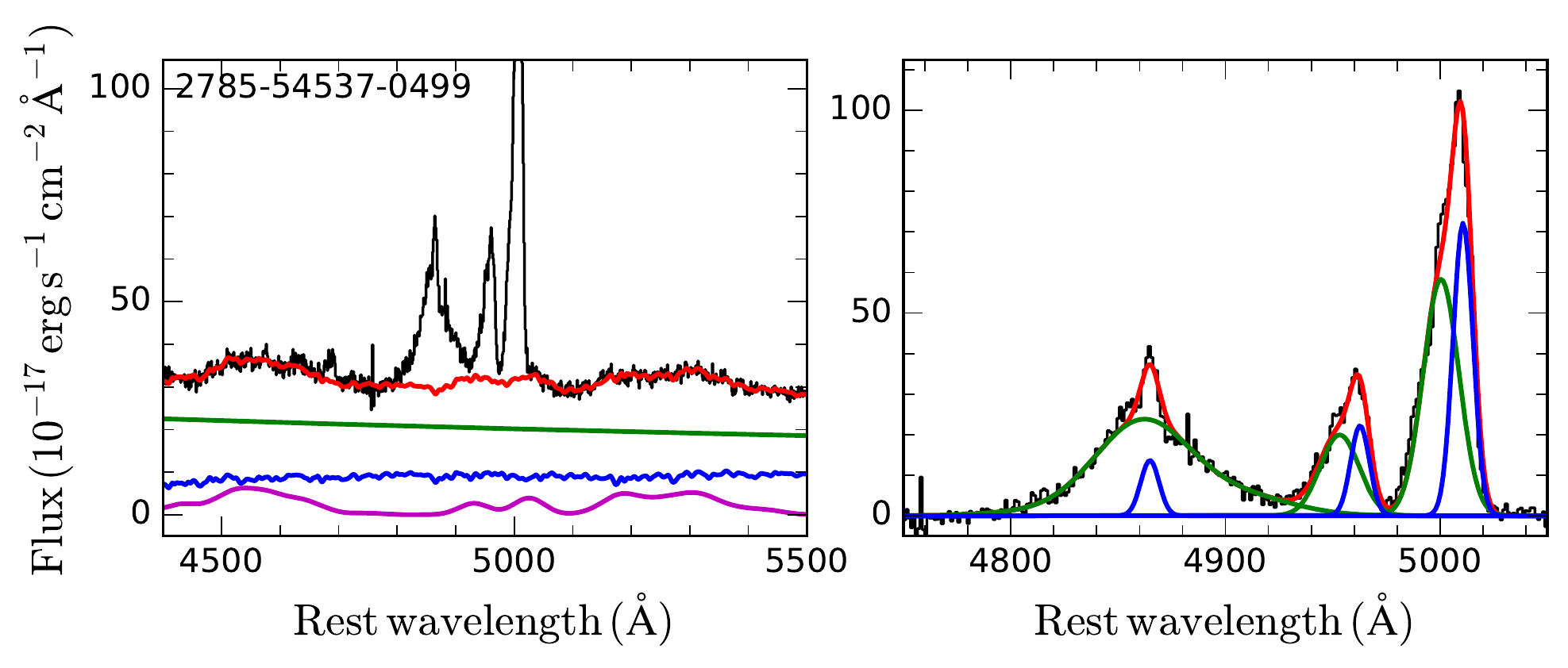}}
\resizebox{9cm}{3.5cm}{\includegraphics{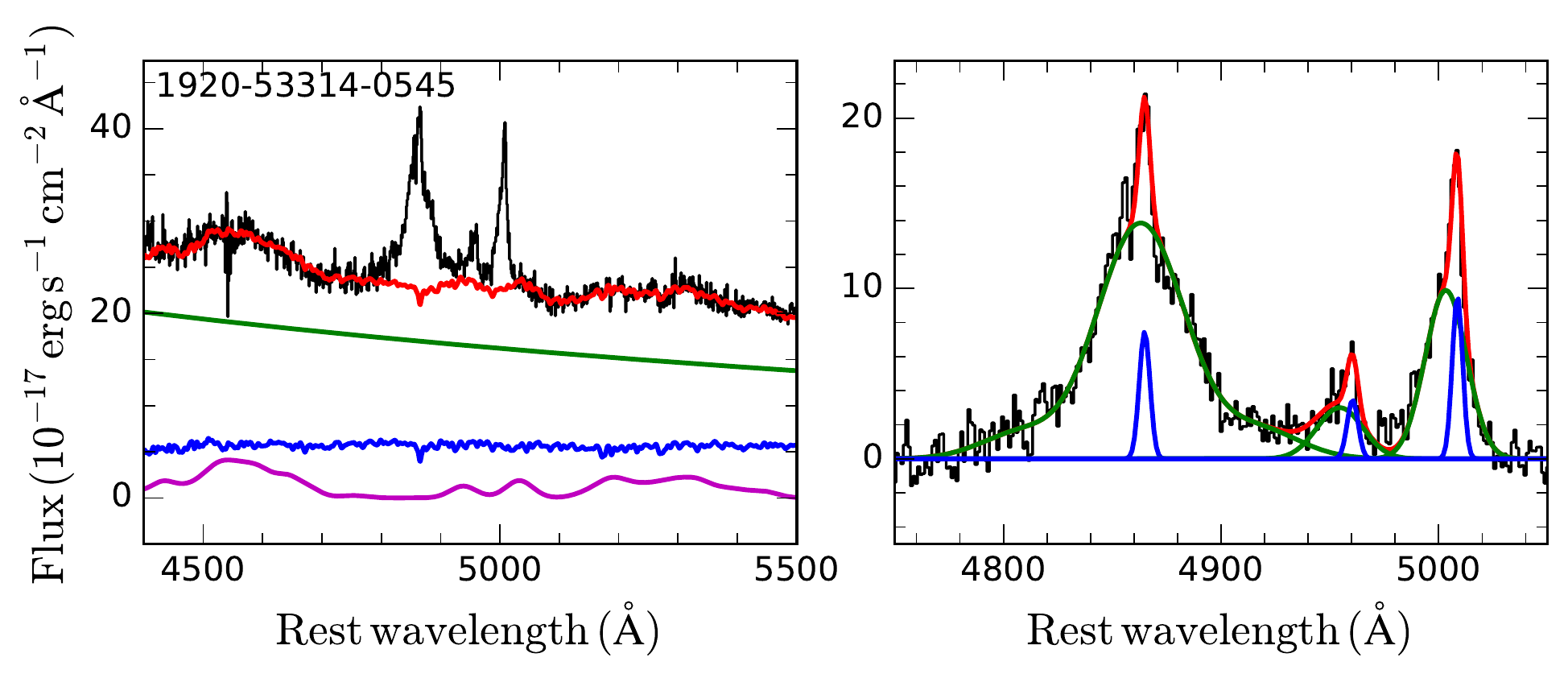}}
\resizebox{9cm}{3.5cm}{\includegraphics{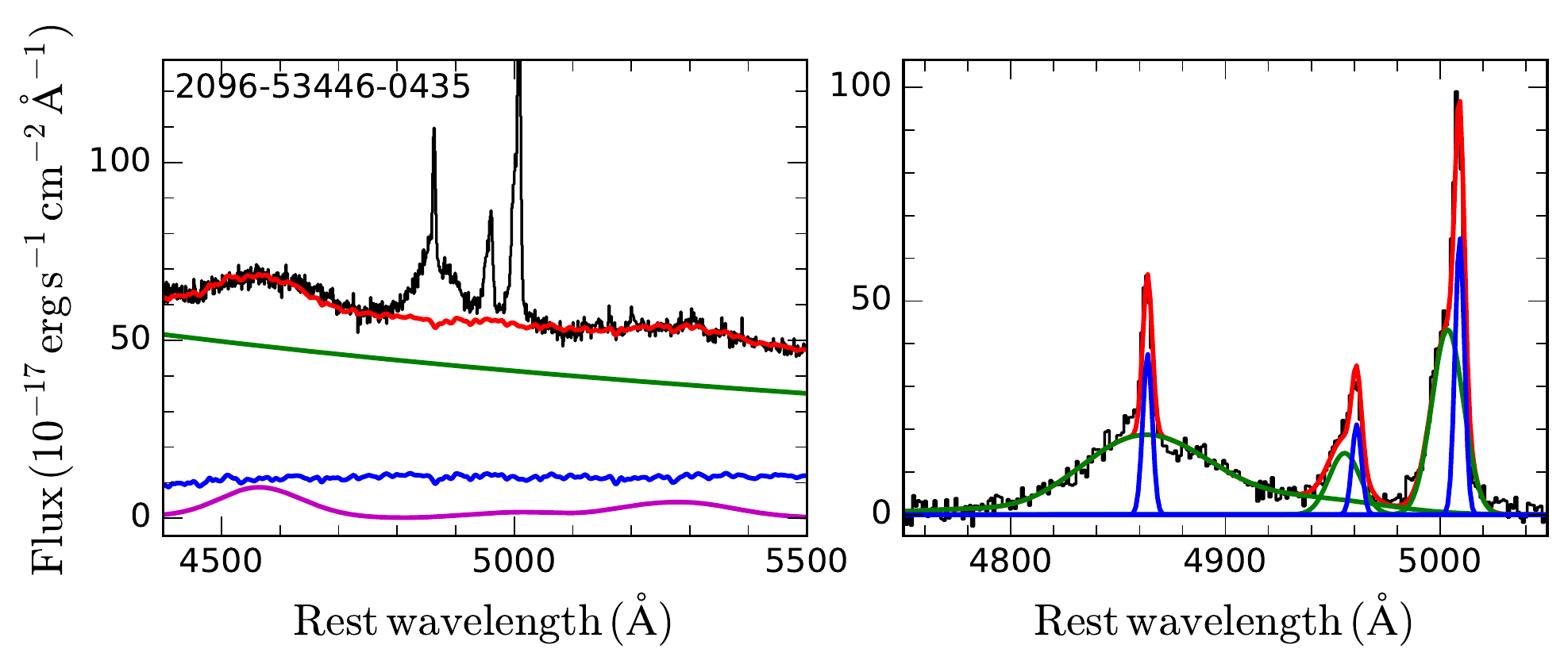}}
\caption{Examples of spectral fitting. Left: continuum modeling. The best-fit model (red), AGN continuum (green), host galaxy contribution (blue), Fe II emission (magenta) are compared with the observed spectrum (black). Right: Emission line modeling. The continuum subtracted spectrum (black) is compared with the best-fit model (red), which is composed of the broad (green) and narrow H$\beta$ (blue), core (blue) and wing (green) components of [O III] doublets. The SDSS ID (Plate-mjd-fiber) of each object is labeled in the left panels.}\label{Fig:spec_fit} 
\end{figure}

To measure the kinematics of outflows, we fitted the [O III] line profile with a double Gaussian model only if the wing (2nd Gaussian) component has A/N $>3$, otherwise, a single Gaussian profile was adopted as the [O III] profile, as similarly done for Type 2 AGNs. The flux, flux weighted center (1st moment), and line dispersion (2nd moment, $\sigma_{\mathrm{line}}$) were calculated from the best-fit model. A few examples of the spectral fitting are shown in Figure \ref{Fig:spec_fit}.

The flux weighted center was calculated from the first moment of the line profile as   
\begin{equation}
\lambda_{\mathrm{avg}}= \frac{\int \lambda f_{\lambda} d\lambda}{\int f_{\lambda} d\lambda},
\end{equation}         
where $f_{\lambda}$ is the best-fit model flux at each wavelength.

\begin{figure}
\centering
\resizebox{9cm}{7.0cm}{\includegraphics{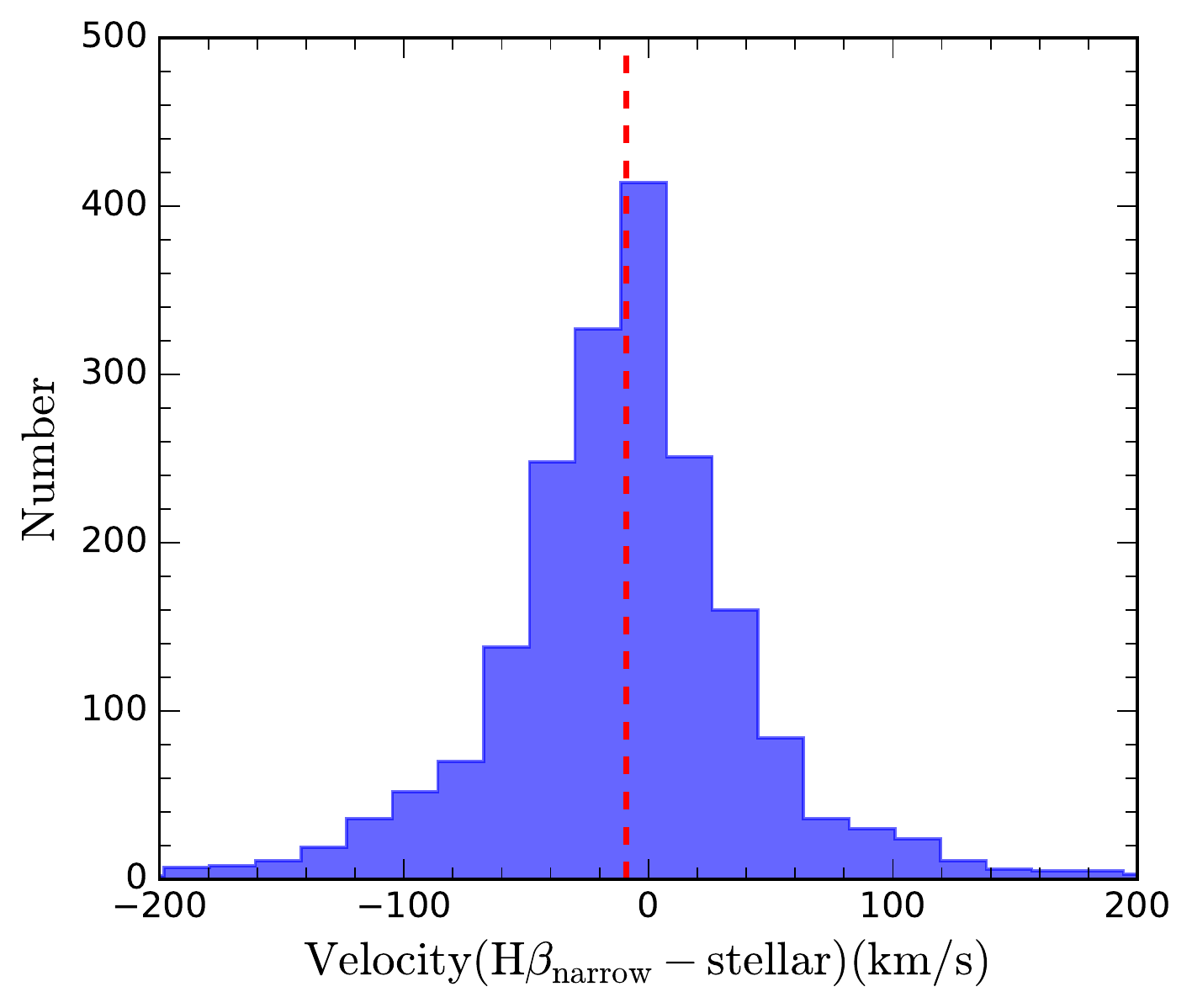}}
\caption{Velocity shift of the H$\beta$ narrow component with respect to systemic velocity based on stellar lines. The dashed line represents the average of the distribution.}\label{Fig:shift} 
\end{figure}

Then, the velocity shift of [O III] was calculated with respect to the systemic velocity, which was measured from the stellar component for 38\% of the sample, in which stellar continuum contribution is greater than 50 \% of the total continuum flux. As mentioned above, the systemic velocity based on stellar absorption lines is difficult to measure in Type 1 AGNs compared to Type 2 AGNs. Instead, the narrow H$\beta$ component is widely used as a reference for the systemic velocity. In Figure \ref{Fig:shift}, we plotted the velocity shift of the H$\beta$ narrow component centroid with respect to stellar absorption lines for those 38\% objects. The narrow component of H$\beta$ shows an average velocity shift of  $-9^{+41}_{-45} \mathrm{km\, s^{-1}}$ (indicated by the red dashed line) with respect to stellar lines, suggesting that the H$\beta$ narrow component is a good proxy for stellar absorption lines with somewhat larger uncertainties. Therefore, for the rest of the sample (62\%), for which we could not measure the systemic velocity from stellar absorption lines, we used the narrow H$\beta$ component as a reference for the systemic velocity.

The flux weighted second moment was used to represent the velocity dispersion of [O III] as
\begin{equation}
\sigma_{\mathrm{line}}= \frac{\int {\lambda}^2 f_{\lambda} d\lambda}{\int f_{\lambda} d\lambda} - {\lambda}^2_{\mathrm{avg}}
\end{equation}         
Uncertainties of all parameters e.g., flux, velocity shift, and velocity dispersion were estimated using Monte Carlo simulations generating 100 mock spectra by adding Gaussian noise to the observed spectrum with the flux uncertainty associated to it. From the distribution of individual parameters having 100 measurements each, we calculated 1$\sigma$ dispersion and considered it as the measurement uncertainty of that parameter. The mean fractional error of $\sigma_{\mathrm{[O\,III]}}$ is $-0.89 \pm 0.33$ in the logarithmic scale, corresponding to $\sim 12$\% uncertainty. The mean uncertainty of $V_{[\mathrm{O III]}}$ is $22.3 \pm 12.8$ km s$^{-1}$, which is $\sim$25\%.           

\begin{figure}
\centering
\resizebox{9cm}{8.0cm}{\includegraphics{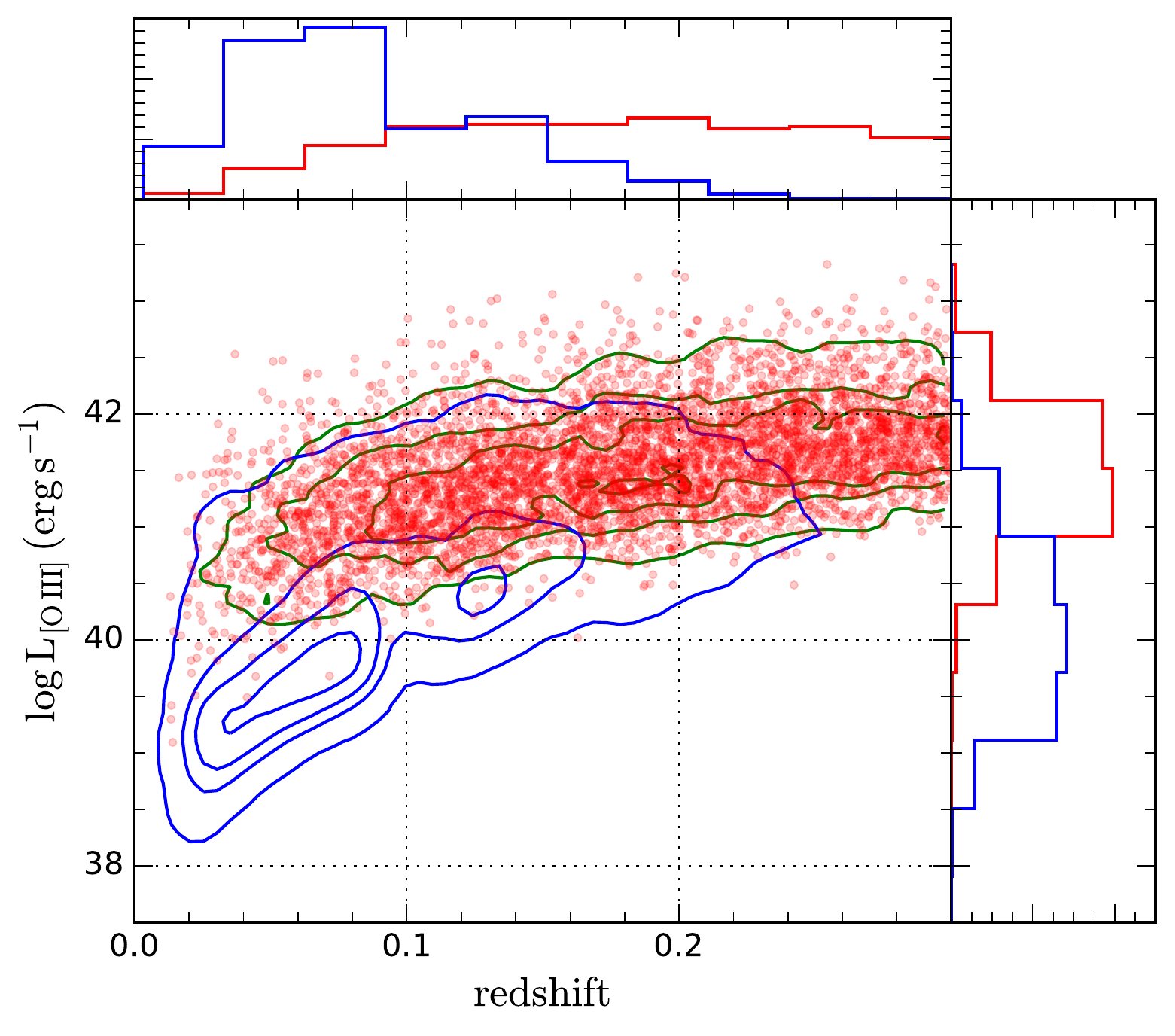}}
\caption{Distribution of [O III] luminosity ($L_{[\mathrm{O III]}}$) and redshift of the Type 1 AGN sample (red). The 25, 50, 75 and 99 percentile density contours are shown for Type 1 (green) and Type 2 AGNs (blue). The normalized histograms of redshift (top) and $L_{[\mathrm{O III]}}$ (right) are also shown for Type 1 (red) and Type 2 (blue).}\label{Fig:L_z} 
\end{figure}

The measured line widths are corrected for the instrumental resolution. We note that a number of objects have $\sigma_{\mathrm{[O\,III]}} < 30$ km s$^{-1}$, much smaller than the instrumental resolution, while some objects have a fractional error in $\sigma_{\mathrm{[O\,III]}} >1$, indicating that the measurement is largely uncertain. As performed in the analysis of Type 2 AGNs, we have excluded those objects (i.e., either $\sigma_{\mathrm{[O\,III]}} < 30$ km s$^{-1}$ or the fractional error in $\sigma_{\mathrm{[O\,III]}} >1$), which are 496 objects in total. Thus, we focus on 5221 Type 1 AGNs for the [O III] kinematics study.

We combined the aforementioned Type 1 AGN sample with the Type 2 AGN sample presented by \citet{2016ApJ...817..108W}, which contains $\sim 39,000$ objects at $z<0.3$. The Type 2 AGN sample has a mean uncertainty in $\sigma_{\mathrm{[O\,III]}}$ of $\sim$14 \%, while the mean uncertainty of $V_{[\mathrm{O III]}}$ is $27.5 \pm 14.6$ km s$^{-1}$. We present the distribution of [O III] luminosity (un-corrected for dust extinction) and redshift, respectively for Type 1 and Type 2 AGNs in Figure \ref{Fig:L_z}. Type 1 AGNs are on average more luminous than Type 2 AGNs. The mean of $\log L_{\mathrm{[O\,III]}}$ is $41.5 \pm 0.52$ and $40.1\pm 0.71$, respectively, for Type 1 and Type 2 AGNs, while the mean $z$ is 0.172 and 0.086, respectively, for Type 1 and Type 2 AGNs. Note that the apparent positive relation between $\log L_{\mathrm{[O\,III]}}$ and $z$ is merely due to the selection effect and does not reflect any luminosity evolution as our objects are selected based on the S/N criterion.

\section{Results}\label{sec:results}


\begin{figure*}
\centering
\resizebox{16.0cm}{5.5cm}{\includegraphics{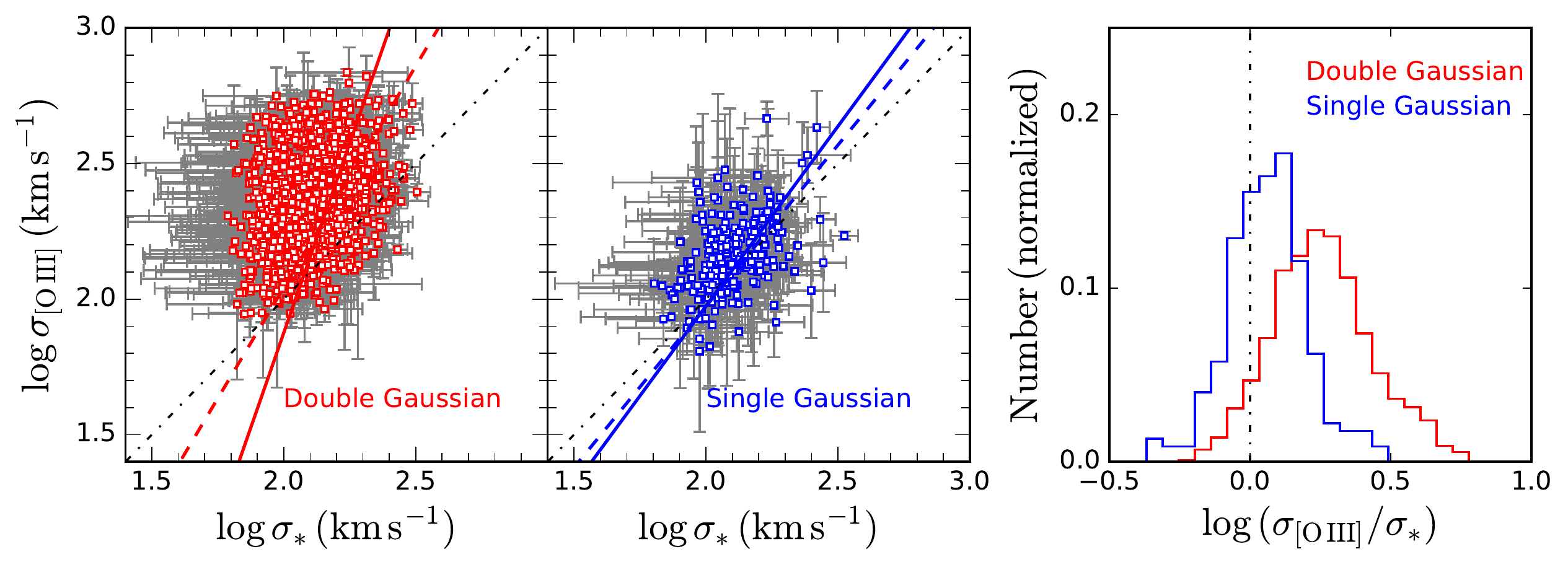}}
\caption{Velocity dispersion of [O III] versus stellar velocity dispersion for Type 1 AGNs: [O III] fitted with a double Gaussian model (left) or a single Gaussian model (middle). The best-fit relation for Type 1 AGNs is represented by solid line while the dashed-dot line denotes the unity relation. The best-fit relation for Type 2 AGNs from \citet{2016ApJ...817..108W} is also shown (dashed lines). Right panel shows the distributions of the ratio of $\sigma_{\mathrm{[O\,III]}}$ to $\sigma_*$ for Type 1 AGNs: double Gaussian [O III] (red) and single Gaussian [O III] (blue).}\label{Fig:disp_ratio}
\end{figure*}

\subsection{Gravitational versus non-gravitational component}
To study the outflow kinematics based on [O III], we first investigate the effect of the host galaxy gravitational potential on the [O III] line profile. We compare velocity dispersion of [O III] with stellar velocity dispersion ($\sigma_*$), which represents the virial motion due to the host galaxy potential. We measured $\sigma_*$ from the stellar template fitting process for $\sim$38\% of Type 1 AGNs in our sample. We fit the spectral region of Mg $b$-Fe covering a wavelength range of $5050 -5400$\AA, using the penalized pixel-fitting ({\sc pPXF}) code \citep{2004PASP..116..138C}. We use MILES stellar spectral library \citep{2010MNRAS.404.1639V} consisting of stellar template of ages 0.06 Gyr to 15.84 Gyr and metallicity [M/H] of -1.71 to 0.2. The $\sigma_*$ measurements are corrected for the SDSS instrumental resolution, which is $\sim$55 km s$^{-1}$ in the spectral range of $5050 -5400$\AA. A majority of these AGNs show a relatively broad [O III] line, which is fitted with a double Gaussian model. Comparing $\sigma_{\mathrm{[O\,III]}}$ with $\sigma_*$ in Figure \ref{Fig:disp_ratio} (left panel), we find that the correlation is weak, indicating the presence of a non-gravitational component in [O III] as similarly found for Type 2 AGNs \citep[][see Figure 4]{2016ApJ...817..108W}. For AGNs with [O III] fitted with a single
Gaussian model, we also compared $\sigma_{\mathrm{[O\,III]}}$ with $\sigma_*$ (middle panel in Figure \ref{Fig:disp_ratio}).

We perform regression analysis after accounting for the measurement uncertainties in both $\sigma_{\mathrm{[O\,III]}}$ and $\sigma_*$. The best-fit for Type 1 AGNs is shown by the solid line in Figure \ref{Fig:disp_ratio}, while for comparison we also plotted the best-fit for Type 2 AGNs from \citet{2016ApJ...817..108W} (dashed line). Interestingly, when the wing component is present in [O III] (left panel), the best-fit deviates from the unity (dashed-dot line), having a steeper slope of $2.79 \pm 0.14$ for Type 1 AGNs and $1.66 \pm 0.01$ for Type 2 AGNs \citep[see][for Type 2 AGNs]{2016ApJ...817..108W}. However, when the wing component is absent (middle panel), the best-fit line has a slope of $1.32\pm 0.11$ for Type 1 and $1.18\pm 0.01$ for Type 2 AGNs much closer to the unity relation. We also present the distribution of the velocity dispersion ratio of [O III] to $\sigma_*$ (right panel in Figure \ref{Fig:disp_ratio}). While [O III] fitted with a double Gaussian is much broader than stellar lines (red histogram), for many objects [O III] fitted with single Gaussian (blue histogram) is also broader than stellar lines, suggesting that outflow signature is significantly present. Considering [O III] dispersion is a combined effect of gravitational potential ($\sigma_{\mathrm{gr}}$) and non-gravitational effect e.g., outflow components ($\sigma_{\mathrm{non-gr}}$), the measured velocity dispersion can be expressed as:
\begin{equation}
\sigma_{\mathrm{total}}=\sqrt{(\sigma_{\mathrm{gr}})^2 + (\sigma_{\mathrm{non-gr}})^2}.
\end{equation}
Note that although $\sigma_{\mathrm{total}}$ may represent more complex nature of gas kinematics, we simply assume that the line-of-sight velocity distribution of gas manifests the convolution of gravitational and non-gravitational components. We expect $\sigma_{\mathrm{[O\,III]}}$ to be a factor of 1.4 larger than $\sigma_{\mathrm{gr}}$ if $\sigma_{\mathrm{non-gr}} = \sigma_{\mathrm{gr}}$. Assuming $\sigma_{\mathrm{gr}}=\sigma_*$, we find an error-weighted mean ratio of $\sigma_{\mathrm{total}}$ to $\sigma_{\mathrm{gr}}$ is 1.0 and 2.3, respectively for Type 1 AGNs with single Gaussian [O III] and double Gaussian [O III]. The non-linearity of the $\sigma_{\mathrm{O III}} - \sigma_*$ relation suggests that the effect of the non-gravitational component is significant and outflows are  common phenomena in both Type 1 and Type 2 AGNs.


\subsection{Outflow fractions}

\begin{figure}
\centering
\resizebox{9cm}{4.2cm}{\includegraphics{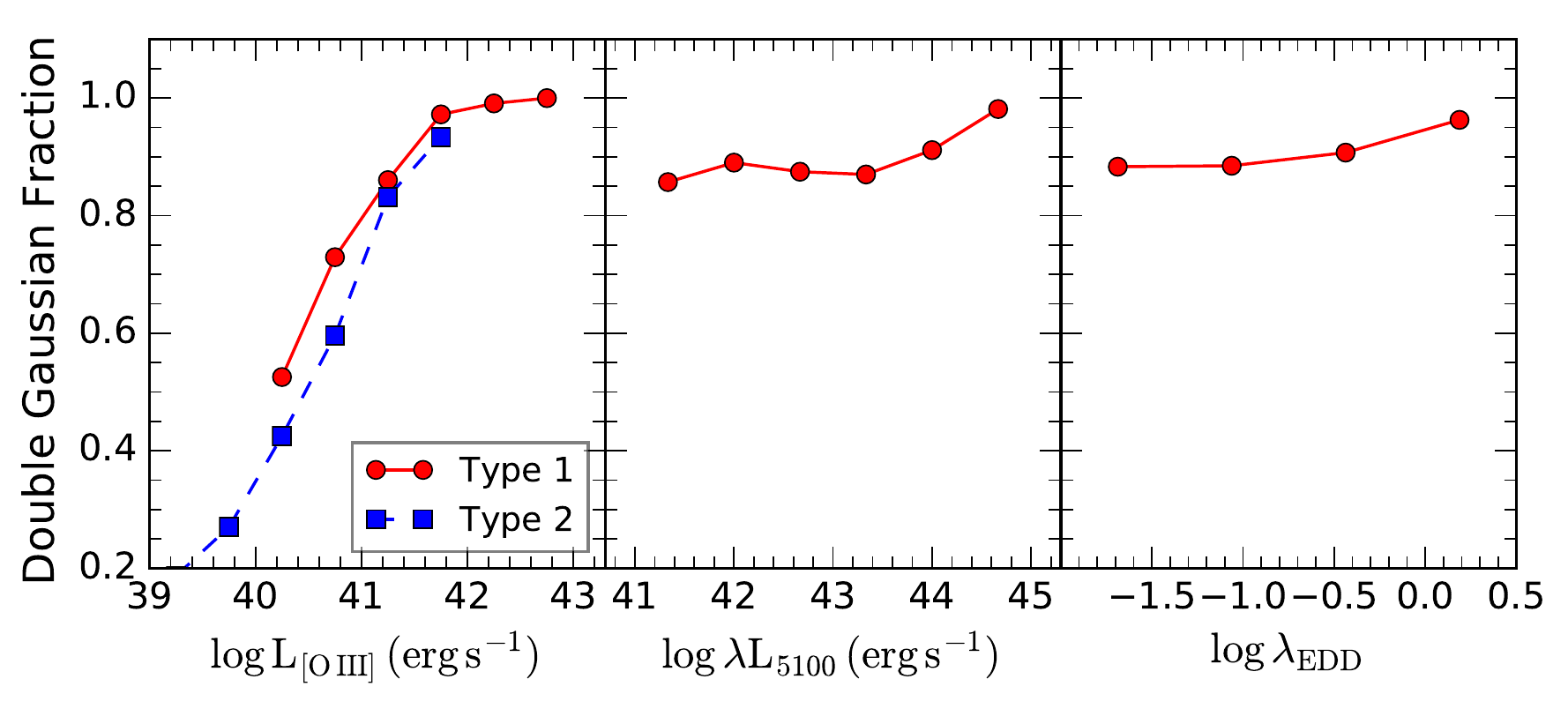}}
\caption{Double Gaussian fraction against [O III] luminosity of Type 1 and Type 2 AGNs (left), double Gaussian fraction against $L_{5100}$ (middle) and Eddington ratio (right) for Type 1 AGNs.}\label{Fig:double_Gaussian_fraction} 
\end{figure}

To study the outflow fractions, we estimated the fraction of AGNs with double Gaussian [O III] in the sample. In the case of Type 1 AGNs, 89\% have double Gaussian [O III] while only 11\% have single Gaussian [O III]. The double Gaussian [O III] fraction in Type 1 AGNs is a factor of 2 larger than that of Type 2 AGNs (43\%). Since the mean luminosity of Type 1 AGNs is much higher than that of Type 2 AGNs, the difference of the double Gaussian [O III] fraction is due to the luminosity effect. 
In Figure \ref{Fig:double_Gaussian_fraction}, we present the fraction of double Gaussian [O III] as a function of [O III] luminosity (left panel). The double Gaussian fraction steeply increases with increasing [O III] luminosity, from $\sim$20\% to $\sim$90\% for Type 2 AGN sample (blue), and from $\sim$40\% to $\sim$100\% for Type 1 AGN sample (red). 
While we find a dramatic increase of the double Gaussian [O III] fraction with [O III] luminosity, the difference of the double Gaussian [O III] fraction is not very large between Type 1 and Type 2 AGNs at a fixed [O III] luminosity.

To better study the effect of AGN luminosity and Eddington ratio on the outflow fraction, we use Type 1 AGNs to calculate the bolometric luminosity and black hole mass. Black hole mass was estimated using the virial relation given in \citet{2015ApJ...801...38W} based on the FWHM of H$\beta$ line and the luminosity at 5100 \AA. 
Then, Eddington luminosity was determined using the relation $L_{\mathrm{EDD}}= 1.26 \times 10^{38} \, M_{\mathrm{bh}}$, while the bolometric luminosity 
is estimated using $L_{\mathrm{bol}} =9 \times L_{5100}$ \citep{2000ApJ...533..631K}. Finally, Eddington ratio is determined by calculating the ratio of bolometric to Eddington luminosity. 
Note that the bolometric luminosity and Eddington ratio of Type 2 AGNs are highly uncertain since both black hole mass and bolometric luminosity are difficult to properly determine. For the Type 2 AGN sample, \citet{2016ApJ...817..108W} estimated black hole mass using  the black hole mass$-$stellar mass relation from \citet{2003ApJ...589L..21M}, and utilized the extinction-uncorrected [O III] luminosity and a scale factor of 3500 from \citet{2004ApJ...613..109H} to estimate the bolometric luminosity.

 In Figure \ref{Fig:double_Gaussian_fraction}, we investigate the double Gaussian fraction as a function of $L_{5100}$ (middle) and Eddington ratio (right) for Type 1 AGNs. Over the large range of optical luminosity and Eddington ratio, the double Gaussian [O III] fraction is at least 80\% and slowly increases to $\sim$100\%, indicating that outflows are commonly detected in Type 1 AGNs.

\begin{figure}
\centering
\resizebox{9cm}{5cm}{\includegraphics{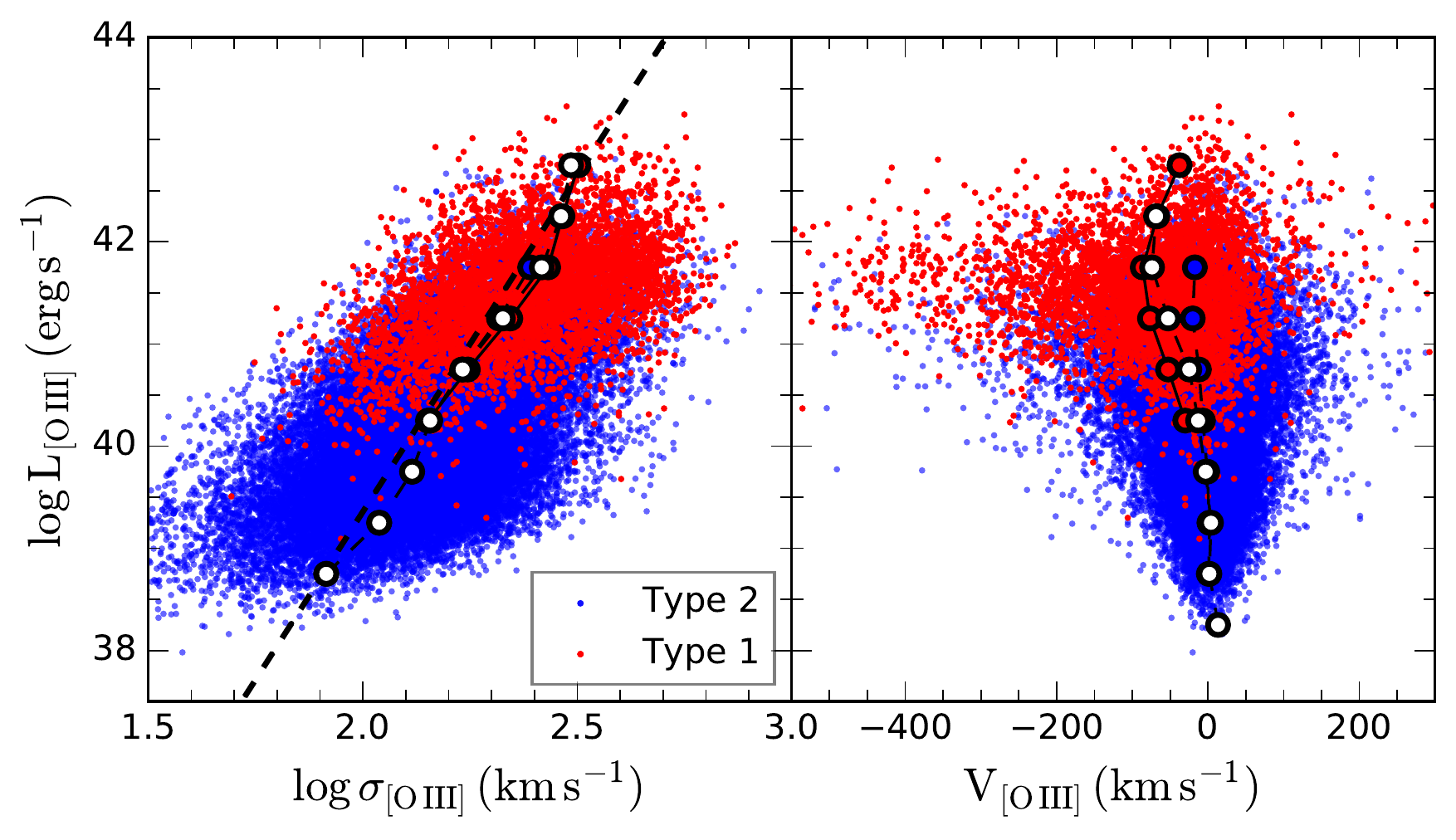}}
\caption{The [O III] luminosity versus velocity dispersion (left panel) and velocity shift (right panel). The Type 1 AGNs are plotted in red while Type 2 AGNs are in blue. The mean values in each $L_{\mathrm{[O\,III]}}$ bins are denoted by red (Type 1 AGNs), blue (for Type 2 AGNs) and white (total sample) circles. The best-fit relation (left panel) for all objects (Type 1 and Type 2 AGNs) including the measurement error is shown by dashed line. On the right panel, the white circles are the mean V$_{\mathrm{[O\,III]}}$ at different $L_{\mathrm{[O\,III]}}$ bins for objects with V$_{\mathrm{[O\,III]}}$ measurement better than 1$\sigma$.}\label{Fig:vvd_LOIII_Ledd}
\end{figure}

\subsection{Velocity shift and velocity dispersion}
We investigate the effect of AGN luminosity on the kinematics of [O III] based on the velocity dispersion and shift of [O III]. First, we investigate [O III] velocity dispersion ($\sigma_{\mathrm{[O\,III]}}$, left panel) against [O III] luminosity for Type 1 (red dots) and Type 2 AGNs (blue dots) in Figure \ref{Fig:vvd_LOIII_Ledd}. A clear positive correlation shows that $\sigma_{\mathrm{[O\,III]}}$ steeply increases with $L_{\mathrm{[O\,III]}}$. 
The mean in each $L_{\mathrm{[O\,III]}}$ bin is plotted for Type 1 (red circles) and Type 2 AGNs (blue circles), which are consistent with the mean of the combined sample (white circles). We perform regression analysis including the measurement errors on both $L_{\mathrm{[O\,III]}}$ and $\sigma_{\mathrm{[O\,III]}}$, finding the best-fit relation for the combined sample (dashed line) as     
\begin{equation}
\log L_{[\mathrm{O III}]} = (6.57 \pm 0.04) \times  \log \sigma_{\mathrm{[O\,III]}} + (26.21 \pm 0.10).
\end{equation}
 A Spearman's correlation test confirms a strong positive correlation between $\log L_{\mathrm{[O\,III]}}$ and $\sigma_{\mathrm{[O\,III]}}$ with a coefficient of 0.44 and 0.52 for Type 1 and Type 2 AGNs, respectively. Note that the correlation remains strong when the sample is divided into different redshift bins, having Spearman's correlation coefficient ($r_s$) of 0.47 ($z<0.1$), 0.59 ($0.1<z<0.2$) and 0.45 ($z>0.2$), thereby eliminating the possibility of any possible selection bias. Such a strong correlation indicates that the effect of the non-gravitational component increases with AGN luminosity.

Second, we investigate velocity shift of [O III] (V$_{\mathrm{[O\,III]}}$) against $L_{\mathrm{[O\,III]}}$ (right panel in Figure \ref{Fig:vvd_LOIII_Ledd}). Interestingly, the number of AGNs with blueshifted [O III] is much larger than that of AGNs with the redshifted [O III]. The number ratio of blueshifted  to redshifted [O III] ($N_\mathrm{BR}$) is found to be 3.6, which is much larger compared to Type 2 AGNs, which has $N_\mathrm{BR}$= 1.08. We have also calculated $N_\mathrm{BR}$ after eliminating objects with very small V$_{\mathrm{[O\,III]}}$, which are less reliable since the V$_{\mathrm{[O\,III]}}$ has a mean uncertainty (see section \ref{sec:data}) of $\sim 22.3 \pm 12.8$ km s$^{-1}$ for Type 1 and $27.5 \pm 14.6$ km s$^{-1}$ for Type 2 AGNs \citep[see][]{2016ApJ...817..108W}. When reliable measurement of V$_{\mathrm{[O\,III]}}$ better than $1\sigma$ was considered $N_\mathrm{BR}$ is found to be increased to 4.8 for Type 1 and 1.08 for Type 2 AGNs, respectively, and further increased to 6.4 and 1.8 when V$_{\mathrm{[O\,III]}}$ measurements better than $3\sigma$ are exclusively included. This indicates that blueshifted [O III] line is more common than redshifted [O III] in general, while for Type 1 AGNs, blueshifted [O III] is dominant. We interpret this as the effect of orientation as discussed by \citet{2016ApJ...828...97B}. The shift of the [O III] line is mainly caused by the partial extinction due to the dusty stellar disk. For Type 1 AGNs, [O III] is more likely to be blueshifted since the direction of the bicone is close to the line-of-sight and the receding cone is expected to be more easily obscured (for more details, see section \ref{see:vvd}).

\begin{figure}
\centering
\resizebox{9cm}{8cm}{\includegraphics{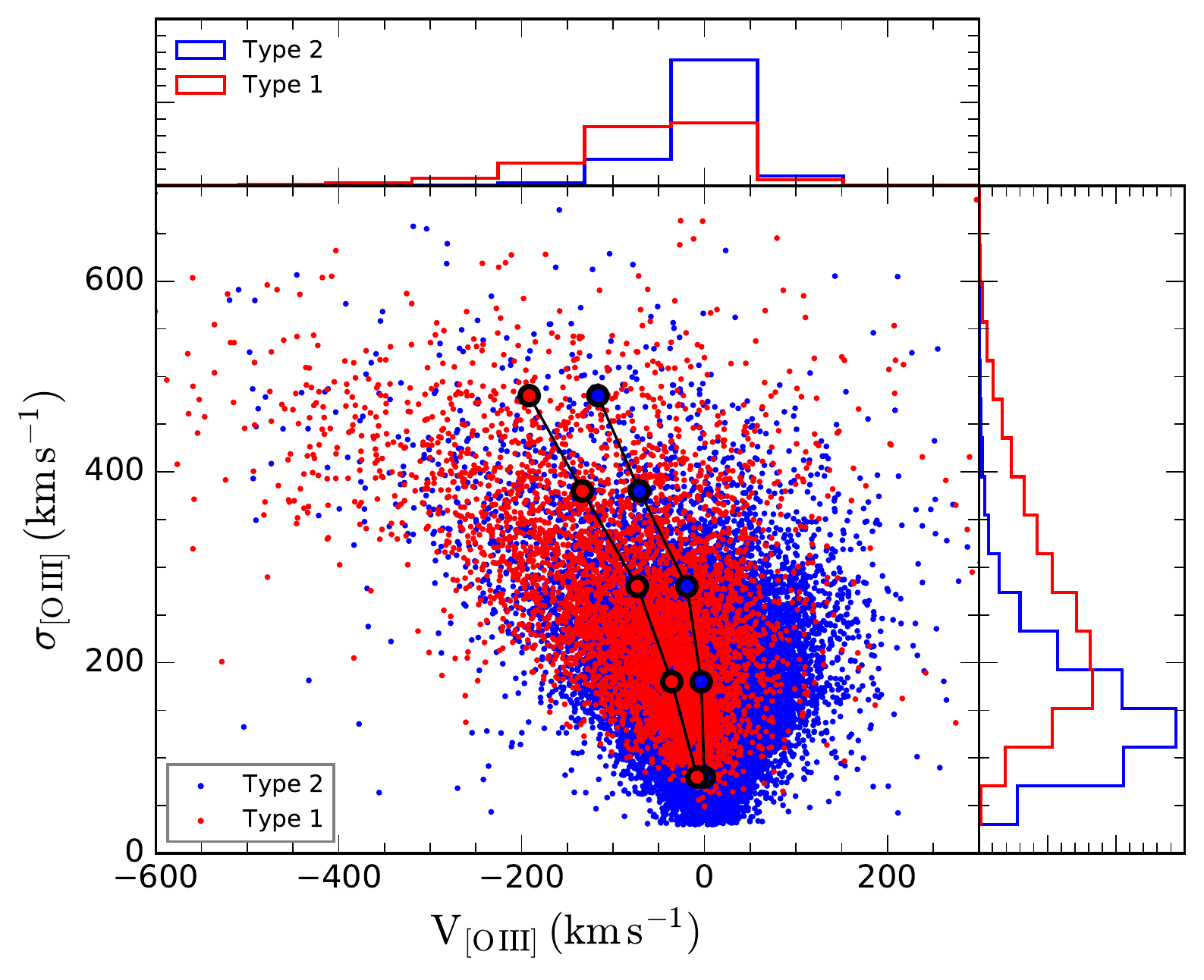}}
\caption{ The [O III] velocity-velocity dispersion (VVD) diagram of AGNs. The Type 1 AGNs are plotted in red while Type 2 AGNs are plotted in blue. The mean velocity shift in each $\sigma_{\mathrm{[O\,III]}}$ bin is denoted by red (for Type 1 AGNs) and blue (Type 2 AGNs) circles. The normalized histograms of $V_{\mathrm{[O\,III]}}$ and $\sigma_{\mathrm{[O\,III]}}$ are shown in the top and right panels for Type 1 (red) and Type 2 AGNs (blue).}\label{Fig:vvd_all}
\end{figure}

\begin{figure*}
\centering
\resizebox{16cm}{4cm}{\includegraphics{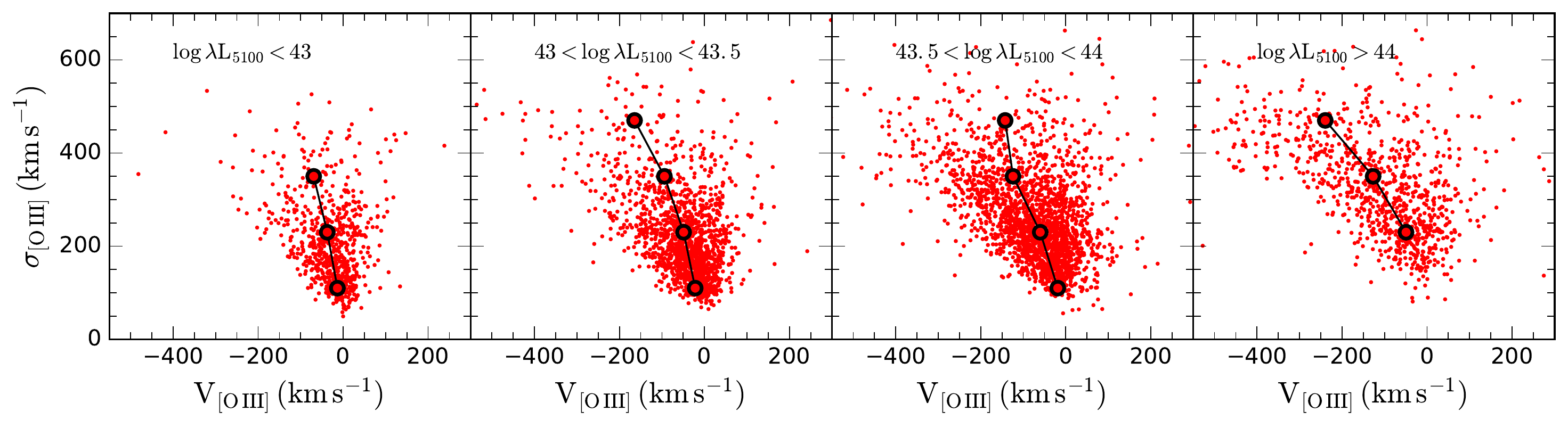}}
\resizebox{16cm}{4cm}{\includegraphics{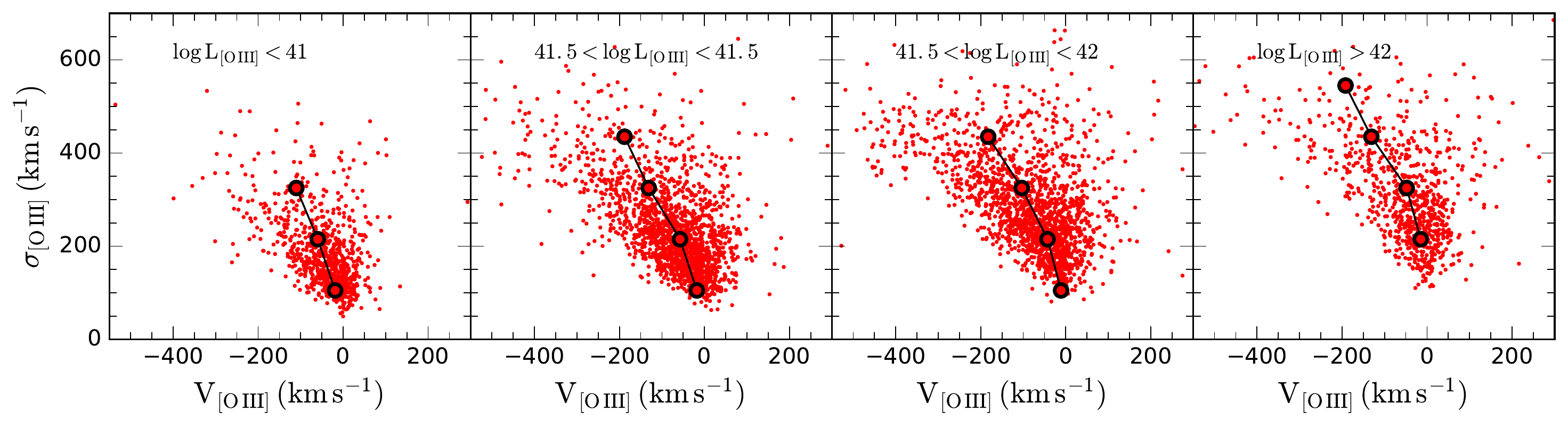}}
\resizebox{16cm}{4cm}{\includegraphics{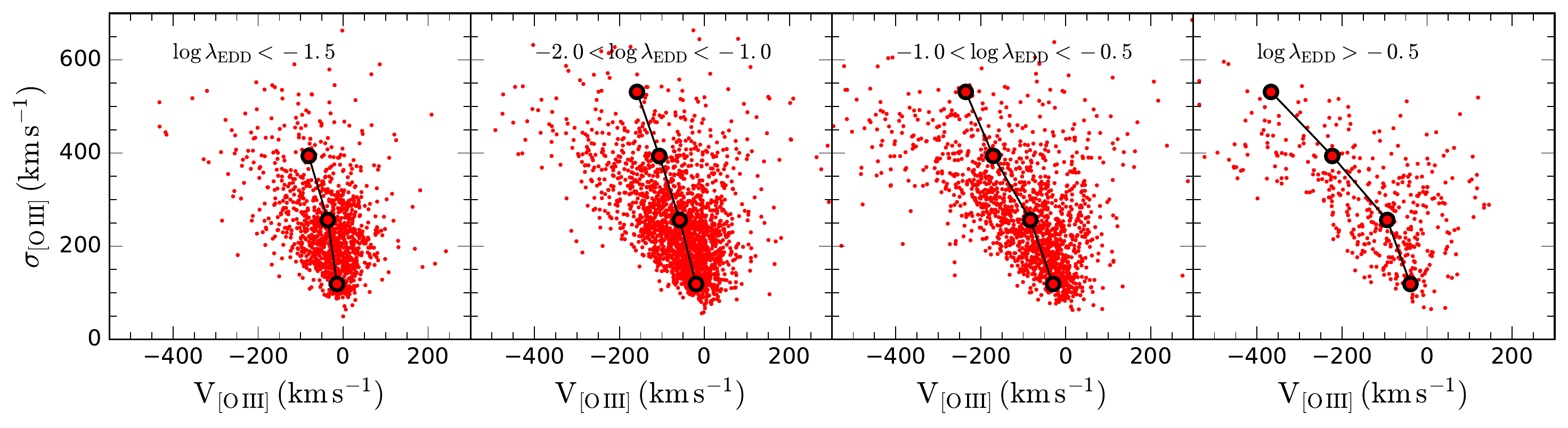}}
\caption{ The VVD diagram of Type 1 AGNs for different $\lambda L_{5100}$ (top), $L_{\mathrm{[O\,III]}}$ (middle) and Eddington ratio (bottom) estimated based on $\lambda L_{5100}$ (see text for more explanation) bins. The mean velocity shift in each $\sigma_{\mathrm{[O\,III]}}$ bin is shown by red circles.}
\label{Fig:vvd_LOIII_bins}
\end{figure*}

We find a slight increase of V$_{\mathrm{[O\,III]}}$ with increasing $L_{\mathrm{[O\,III]}}$ (right panel in Figure \ref{Fig:vvd_LOIII_Ledd}). The mean V$_{\mathrm{[O\,III]}}$ at different $L_{\mathrm{[O\, III]}}$ bins are plotted for Type 1 (red circle) and Type 2 (blue circles) AGNs, which shows a slightly increasing trend with luminosity. However, this trend is much weaker than that of $\sigma_{\mathrm{[O\,III]}}$. While we see that the range of V$_{\mathrm{[O\,III]}}$ is increasing with [O III] luminosity, there are also many AGNs with V$_{\mathrm{[O\,III]}}$ close to zero within the uncertainty. Considering the measurement uncertainty of V$_{\mathrm{[O\,III]}}$, we calculate the mean V$_{\mathrm{[O\,III]}}$ at different $L_{\mathrm{[O\,III]}}$ bins using the AGNs with V$_{\mathrm{[O\,III]}}$ measurement better than 1$\sigma$ (white circles in the right panel). We found a stronger trend of increasing V$_{\mathrm{[O\,III]}}$ with $L_{\mathrm{[O\,III]}}$ i.e. high-luminosity AGNs tend to show more blueshifted [O III], suggesting gas outflows are associated with AGN accretion.


\subsection{The VVD diagram of AGNs}\label{see:vvd}

By combining velocity shift and velocity dispersion of [O III], we investigate the velocity - velocity dispersion (VVD) diagram for Type 1 AGNs (red dots) and compare with the VVD distribution of Type 2 AGNs from \citet{2016ApJ...817..108W} in Figure \ref{Fig:vvd_all}. Type 1 AGNs also show a V-shape structure as Type 2 AGNs. The mean velocity shift at fixed different velocity dispersion bins is shown, respectively for Type 1 (red circle) and Type 2 (blue circles) AGNs. On average, Type 1 AGNs have higher velocity shift (top histogram) and dispersion (right histogram) than Type 2 AGNs. The mean velocity shift and mean velocity dispersion are $-73$ km s$^{-1}$ ($-6$  km s$^{-1}$) and 259 km s$^{-1}$ (156 km s$^{-1}$) for Type 1 (Type 2) AGNs, respectively. Moreover, a majority of Type 1 AGNs are found to have higher blueshift compared to the Type 2 AGNs. Furthermore, AGNs having large [O III] line width are found to be more blueshifted. Note that there is a lack of objects having higher velocity shift and lower velocity dispersion in both Type 1 and 2 AGNs. This indicates that highly blueshifted [O III] tends to have high velocity dispersion. For Type 2 AGNs, \citet{2016ApJ...817..108W} postulated that such effect is due to the intrinsic launching velocity, which increases the kinematic signature making both velocity shift and dispersion larger. This could also be the reason for Type 1 AGNs of lacking high velocity shift at low velocity dispersion. In contrast, at any given velocity dispersion there is a range of velocity shifts. In fact, the majority of the AGNs have very low velocity shift but large dispersion. This could be explained in terms of a biconical outflow model, in which on one hand, the approaching and receding cones are symmetric and the blueshift and redshift cancel each other in the flux-weighted spectra, resulting nearly zero velocity shift. In contrast, the velocity dispersion reflects the Doppler broadening due to the combined effect of blueshift and redshift components. On the other hand, if the dust extinction is significant in some AGNs, the flux from the receding cone (i.e., behind the dusty stellar disk) is reduced, resulting a large blueshifted [O III], while the wide opening angle of the cone is presumably responsible for the large Doppler broadening \citep[see the discussion by][]{2016ApJ...817..108W, 2016ApJ...828...97B}.

In Figure \ref{Fig:vvd_LOIII_bins} we present the VVD diagrams for Type 1 AGNs at different $L_{5100}$ (top), [O III] luminosity (middle) and Eddington ratio (bottom). The mean velocity at fixed $\sigma_{\mathrm{[O\,III]}}$ bins is denoted in each panel by red circles. First, we clearly see an increasing trend of velocity shift and dispersion with $L_{5100}$ and $L_{\mathrm{[O\,III]}}$. While for low-luminosity AGNs the distribution of the VVD is limited to relatively small values of velocity
shift and dispersion, the VVD diagram expands toward large values for higher luminosity AGNs. The fraction of AGNs with $\mathrm{V_{[O\,III]}<-200 \, km \, s^{-1}}$ increases from 2\% to 24\%,  while the fraction of AGNs with $\sigma_{\mathrm{[O\,III]}}>300 \, \mathrm{km \, s^{-1}}$ increases from 15\% to 58\% as luminosity increases from $L_{\mathrm{5100}} <43 \, \mathrm{erg \,s^{-1}}$ to $L_{\mathrm{5100}} > 44 \, \mathrm{erg \,s^{-1}}$.

Second, velocity shift and dispersion are on average increasing with Eddington ratio. At $\log \lambda_{\mathrm{EDD}}<-1.5$, only $\sim 3$\% of AGN shows $\mathrm{V_{[O\,III]}<-200 \, km \, s^{-1}}$, while it increases to 29\% for $\log \lambda_{\mathrm{EDD}}>-0.5$. Similarly, at $\log \lambda_{\mathrm{EDD}}<-1.5$, only 21\% AGNs show $\sigma_{\mathrm{[O\,III]}}>300 \, \mathrm{km \, s^{-1}}$ but the fraction increases to 48\% at $\log \lambda_{\mathrm{EDD}}>-0.5$. We find the same trend that the mean value of velocity shift and velocity dispersion increases with Eddington ratio. These results clearly demonstrate the connection of AGN power with outflows, suggesting that ionized gas outflows are mainly driven by AGNs.

The VVD diagram of Type 1 AGNs as well as that of Type 2 AGNs can be well explained by the 3D biconical outflow model, which was developed by \citet{2016ApJ...828...97B}. Here, we investigated the physical parameters of the outflows based on this model. The bicone model is made of two identical axisymmetric cones whose vertex is located at the center of the bicone. The cones have an inner hollow region parameterized by an inner half-opening angle $\theta_{\mathrm{in}}$ and extended to the outer half opening angle of $\theta_{\mathrm{out}}$. A thin dust plan having extinction of $0-100$\% is added to represent the dusty stellar disk. Furthermore, the dust plane and bicone have independent orientation defined by $i_{\mathrm{dust}}$ and $i_{\mathrm{bicone}}$ ( $i_{\mathrm{bicone}}=0$ means bicone axis is parallel to the sky plane  and the line-of-sight velocity is zero), respectively. For a fixed $i_{\mathrm{dust}}$, as $i_{\mathrm{bicone}}$ increases from zero, the approaching and receding cones do not cancel each other since the dusty stellar disk
preferentially obscures the part of the cone behind the dusty stellar disk, resulting in a blueshifted or redshifted [O III].

 \begin{figure}
 \centering
 \resizebox{9cm}{8cm}{\includegraphics{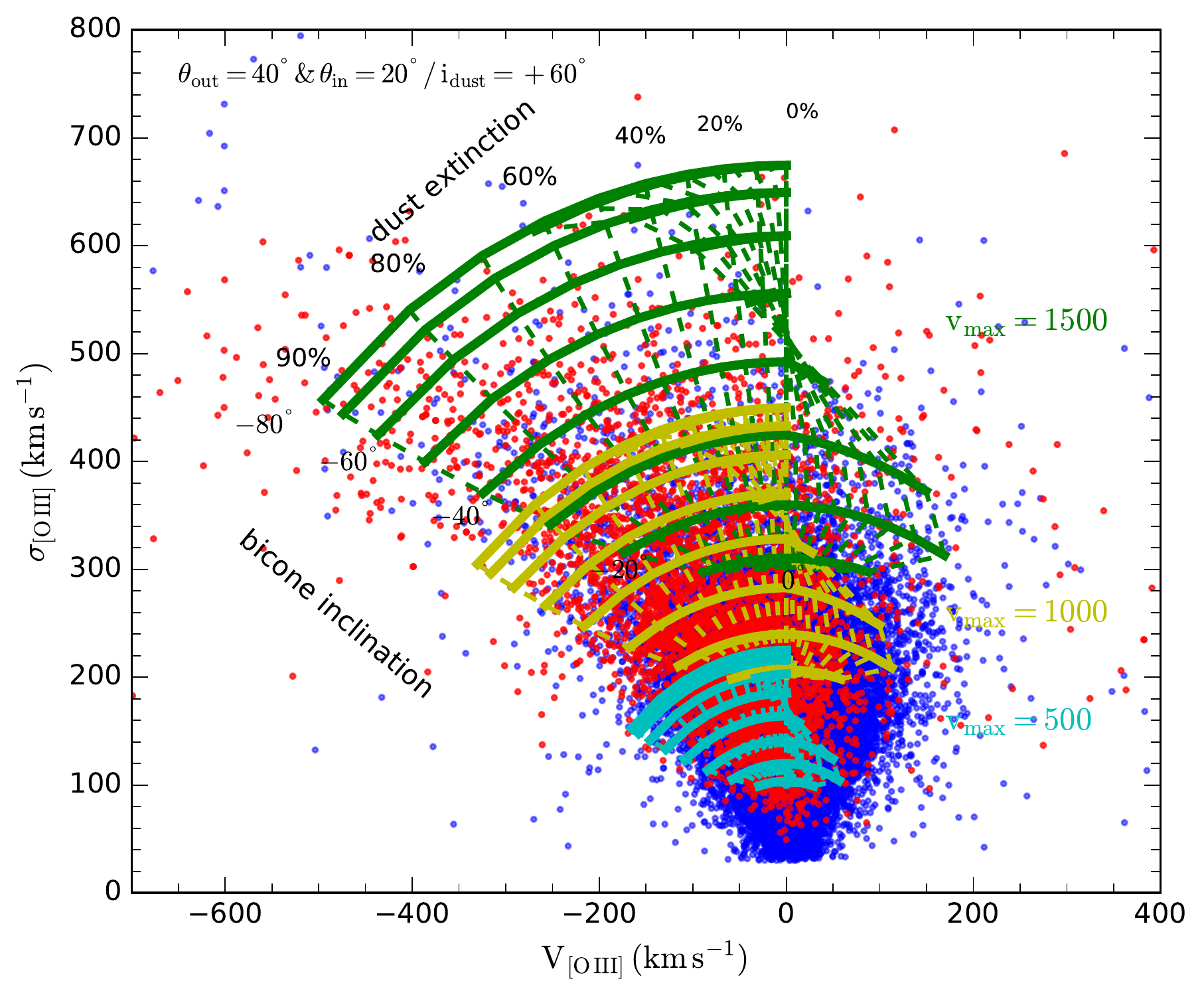}}
 \caption{Model grids (lines) of the biconical outflow model developed by \citet{2016ApJ...828...97B} is overplotted on the VVD diagram for Type 1 (red dots) and Type 2 (blue dots) AGNs. See text for explanations.}\label{Fig:model}
 \end{figure}
 
Based on the Monte Carlo simulations, \citet{2016ApJ...828...97B} found that a wider bicone opening angle increases the number of AGNs with blueshifted [O III]. For Type 1 AGNs in our sample we found that the number ratio of blueshifted to redshifted AGNs is $\sim 3.6$ which can be roughly reproduced by a model with $\theta_{\mathrm{out}}=40\degree$ and $\theta_{\mathrm{in}}=20\degree$. A larger opening angle produces a higher ratio of blueshifted to redshifted AGNs.
For example, the number ratio is $\sim$7 when the half-opening angle $\theta_{\mathrm{out}}$ is 60$\degree$.
In Figure \ref{Fig:model}, we overlaid the model grids, the details of which are presented in \citet{2016ApJ...828...97B}, along with the observed VVD distribution of Type 1 (red dots) and Type 2 (blue dots) AGNs. Here, we showed the model grids with a fixed dust inclination $i_{\mathrm{dust}}=60\degree$ along with 3 sets of different launching velocities ($\mathrm{v_{max}}=$ 500, 1000 and 1500 km s$^{-1}$). For each set, different inclination angles spanning from $-80\degree$ to $+80\degree$ (typical inclination for Type 2 AGNs ranging from $-40\degree$ to $+40\degree$) are presented with a step of $10\degree$, while different dust extinctions spanning between 0\% to 90\% are also plotted with a step of 10\% for illustration. As \citet{2016ApJ...828...97B} demonstrated, velocity dispersion is insensitive to the $i_{\mathrm{dust}}$. Note that AGNs having dust extinction greater than 90\% is rare as the velocity range in the observed VVD diagram can be well produced by dust extinction less than 90\%. 
 
 The intrinsic velocity of outflow ($\mathrm{v_{max}}$) increases $\sigma_{\mathrm{[O\,III]}}$ due to the Doppler broadening. While the AGNs with extreme outflows can have $\mathrm{v_{max}}$ $\sim 1500$ km s$^{-1}$, a majority of low-luminosity AGNs have less significant outflows with $\mathrm{v_{max}} \sim 500$ km s$^{-1}$. At a given $\mathrm{v_{max}}$, the main parameters responsible for the observed integrated gas kinematics are (i) dust extinction and (ii) bicone inclination. If extinction is negligible, V$_{\mathrm{[O\,III]}}$ is independent of $i_{\mathrm{bicone}}$ and remains zero since the velocities of approaching and receding cones cancel each other irrespective of the bicone geometry. In the case of velocity dispersion, $\sigma_{\mathrm{[O\,III]}}$ has the lowest value when the bicone inclination is zero (i.e., perpendicular to the line-of-sight) since the width of the projected velocity distribution is narrowest, while it increases with $i_{\mathrm{bicone}}$. Many AGNs in our sample (both for Type 1 and Type 2) have velocity shift of $\sim$ zero, and velocity dispersion $\sim 200$ $\mathrm{km\,s^{-1}}$ suggesting that they have very low dust extinction. 
 
When dust extinction is significant, V$_{\mathrm{[O\,III]}}$ is highly sensitive to $i_{\mathrm{bicone}}$. As the bicone is more inclined from $0\degree$ to $-80\degree$ (i.e., from Type 2 to Type 1 AGNs), dust obscures a larger part of the receding cone, hence the velocity measured from the flux weighted spectra becomes more negative. In the case of a fixed $i_{\mathrm{bicone}}$, [O III] becomes more blueshifted if the level of extinction increases since the observed velocity distribution becomes more asymmetric. The fact that the number ratio of blueshifted to redshifted AGNs is higher in Type 1 AGNs than Type 2 AGNs can be well explain in terms of the bicone inclination angle. Since Type 1 AGNs have larger bicone inclination angle ($|i_{\mathrm{bicone}}|>40\degree$) than Type 2 AGNs ($|i_{\mathrm{bicone}}|<40\degree$), the former presents larger negative velocity. For example,  V$_{\mathrm{[O\,III]}}$ can be $\sim 1.5$ times more negative at $i_{\mathrm{bicone}}=-90\degree$ (Type 1 AGNs) than $i_{\mathrm{bicone}}=-40\degree$ (Type 2 AGNs) if dust extinction is significant \citep[see][]{2016ApJ...828...97B}. These results are consistent with the orientation effect as expected from the unification model of Type 1 and Type 2 AGNs.

\subsection{Outflow versus radio luminosity}
\begin{figure}
\centering
\resizebox{9cm}{10cm}{\includegraphics{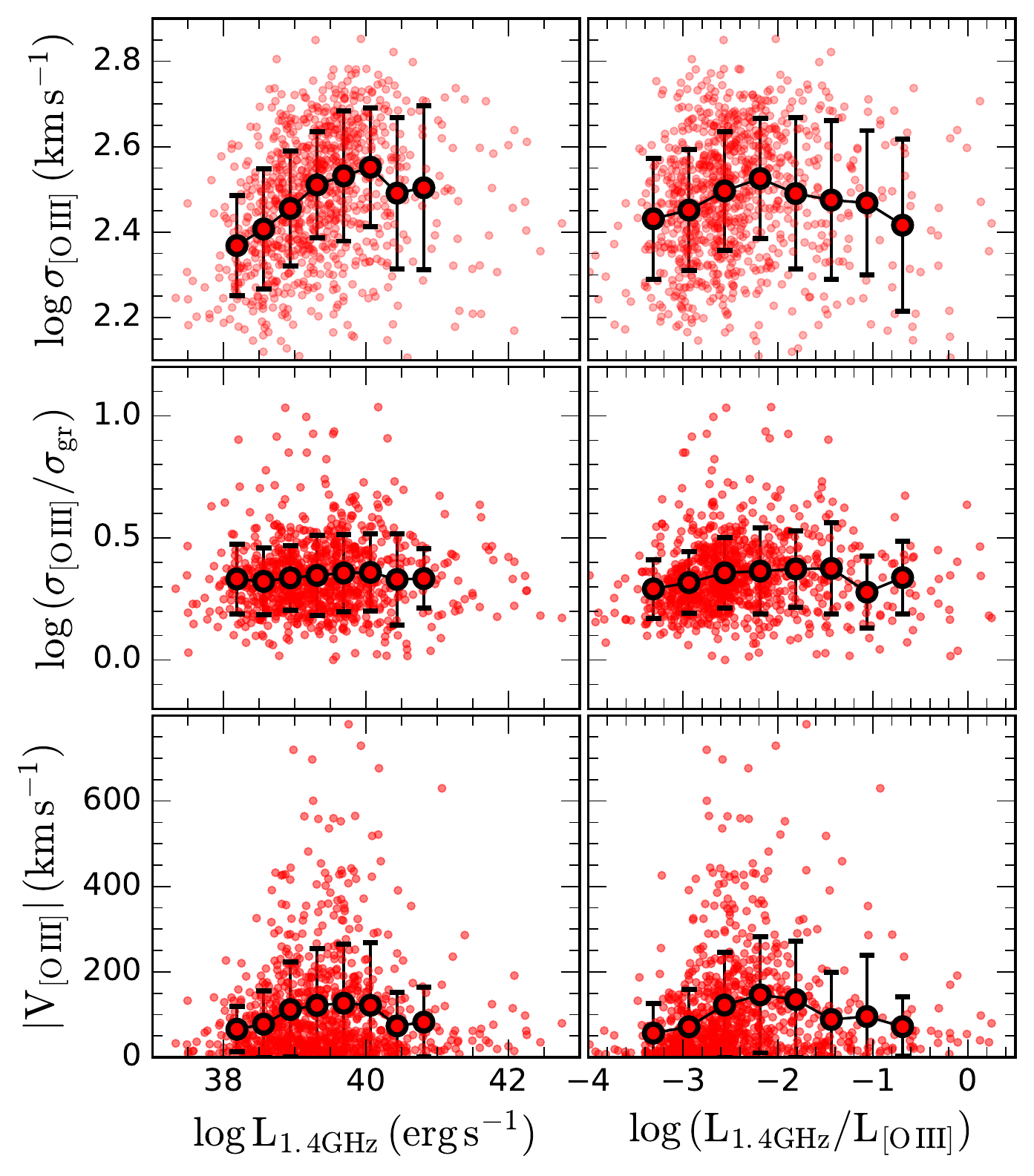}}
\caption{The [O III] velocity dispersion (top), dispersion ratio (middle) and velocity shift (bottom) as a function of luminosity at 1.4GHz (left panels) and the luminosity ratio of 1.4 GHz to [O III] (right panels) for Type 1 AGNs. The mean values at each bins along x-axis are shown along with 1$\sigma$ dispersion. $\sigma_{\mathrm{gr}}=\sigma_{\mathrm{[O\, III], narrow}}$ is considered in the plot.}\label{Fig:radio}
\end{figure}

To study the effect of radio emission on the outflow of AGNs, we created a radio-detected subsample of Type 1 AGNs by cross-matching our Type 1 AGN sample with the VLA FIRST Survey (Catalog version 14dec17\footnote{\url{http://sundog.stsci.edu/first/catalogs.html}}) using a matching radius of $5''$. This search yields 914 Type 1 AGNs corresponding to $\sim 17$\% of the sample. The $\log L_{\mathrm{1.4 GHz}}$ distribution has a mean of $39.3 \pm 0.8$ covering $\sim 4$ orders of magnitude in radio luminosity. In Figure \ref{Fig:radio}, we plotted the [O III] kinematics against radio luminosity ($L_{\mathrm{1.4 GHz}}$, left panels) and radio loudness (defined by $L_{\mathrm{1.4 GHz}}/L_{\mathrm{[O\,III]}}$, right panels). The mean $\sigma_{\mathrm{[O\,III]}}$ in each bin along radio luminosity and radio loudness is plotted to show the average trend (red circle).

The [O III] velocity dispersion is found to increase with radio luminosity (upper-left) and radio loudness (upper-right) as evident from Figure \ref{Fig:radio}. A similar trend has been noticed by \citet{2016ApJ...817..108W} for Type 2 AGNs. However, the authors explained that this apparent trend for Type 2 is due to the correlation between radio luminosity and $\sigma_*$, i.e., the massive galaxies have larger radio luminosity and larger [O III] velocity dispersion. The authors showed for Type 2 AGNs that the trend disappears when [O III] velocity dispersion is normalized by the $\sigma_*$. Since $\sigma_*$ were measured only a fraction of the Type 1 AGNs, we used the dispersion of [O III] core component as a proxy of $\sigma_*$ to study this effect in Type 1 AGNs. We found the $\sigma_{\mathrm{[O\,III]}}/\sigma_{\mathrm{gr}}$ is almost constant (middle panels) across all radio luminosity and radio loudness, thereby suggesting that the kinematics of the non-gravitational component is not due to the radio emission. This is further justified by the plots of [O III] velocity shift versus radio luminosity (lower-left) and radio loudness (lower-right), which also show no correlation with radio activity, especially at high radio luminosity AGNs.

\subsection{Outflow energetics}
To study the energetics of the outflows in Type 1 AGNs, we quantitatively estimated the mass outflow rates, energy injection rates and momentum flux based on a simple outflow model. Here we assumed a biconical outflow and case B recombination \citep{2010ApJ...708..419C,2016ApJ...828...97B}. We note that detailed kinematical modelings of individual objects are necessary for an accurate estimation of these parameters. For a statistical study, we try to determine outflow energetics without spatial information for Type 1 AGN sample. Following \citet{2011MNRAS.415.2359N}, we first estimate the mass of the ionized gas ($M_\mathrm{gas}$) based on the [O\, III] luminosity 
\begin{equation}
M_\mathrm{gas} = 0.4 \times 10^8 M_{\odot} \times (L_{\mathrm{[O\, III],43}}) (100 \, \mathrm{cm^{-3}}/n_e)
\label{eq:mass}
\end{equation} 
where $L_{\mathrm{[O\,III],43}}$ is the total [O\,III] luminosity in the unit of $10^{43}$ erg s$^{-1}$ and $n_e$ is the electron density, which is the largest source of uncertainty in the above equation \citep[see][]{2018arXiv180602839K}. We adopted $n_e$ from \citet{2017ApJS..229...39R}, who measured $n_e$ for a large number of Type 1 AGNs assuming a fixed temperate $T=10^4$ K and the intensity ratio of $\mathrm{[S\, II]\lambda 6716/ \lambda 6731}$ lines for which [S II] doublet fittings were reliable. We found that 1526 objects in our sample have overlapped with the sample of \citet{2017ApJS..229...39R} with a median $n_e$ of 272 cm$^{-3}$. We estimated $M_\mathrm{gas}$ and outflow energetics for these 1526 objects. We caution that $n_e$ is also sensitive to the gas temperature, ionization mechanism and geometry of NLR, a warm low-density gas may contain significant mass and kinematic power but line ratios may not be sensitive to such gas, hence, the estimated $n_e$ from line ratio can be uncertain \citep[see][]{2018NatAs...2..198H}.

\begin{figure}
\centering
\resizebox{9cm}{13cm}{\includegraphics{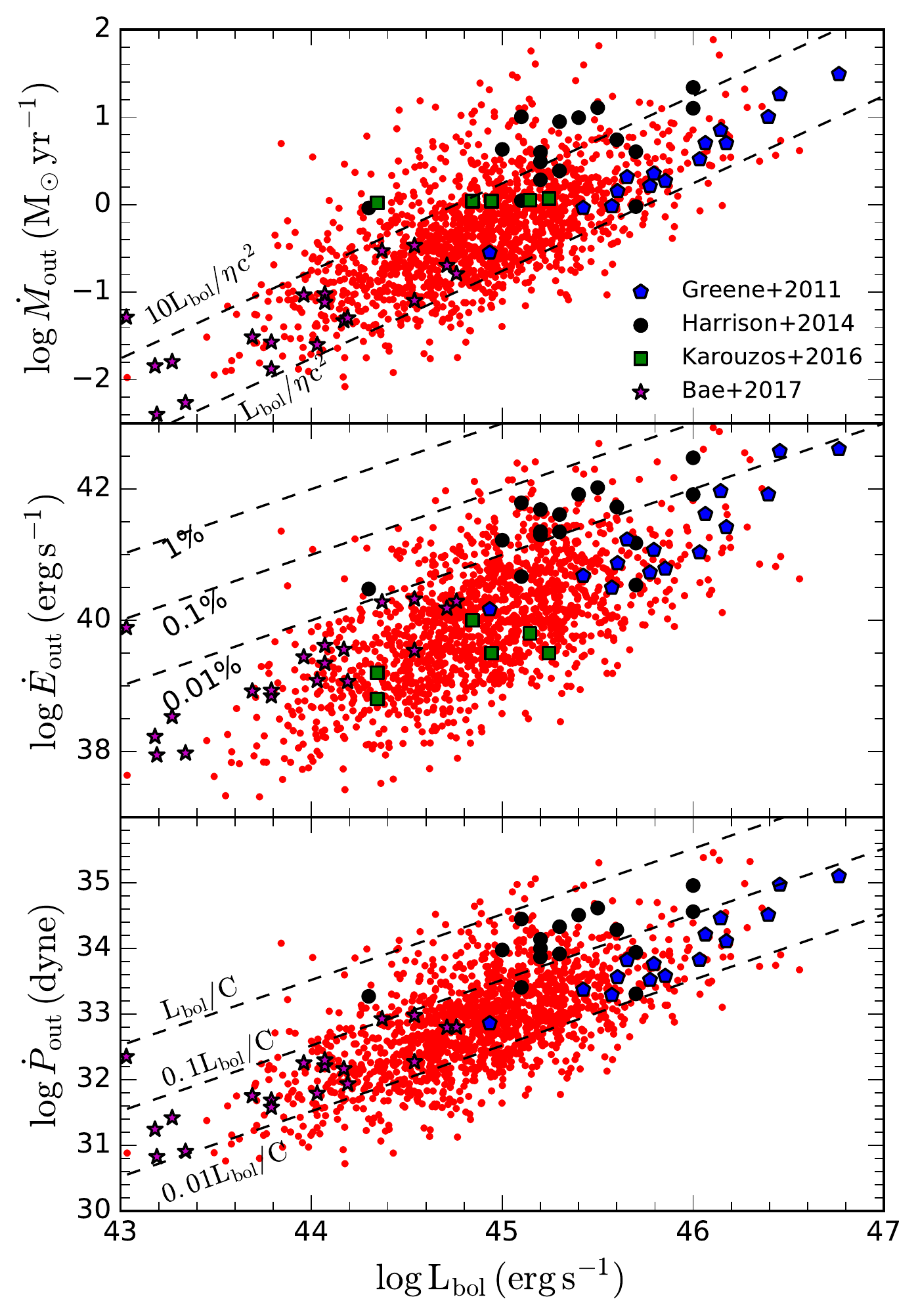}}
\caption{ Outflow energetics versus bolometric luminosity. The mass outflow rate (top), energy injection rate (middle) and momentum flux (bottom) is plotted as a function of bolometric luminosity for Type 1 AGNs. Outflow energetics of some literature sample taken from \citet{2011ApJ...732....9G}, \citet{2014MNRAS.441.3306H}, \citet{2016ApJ...833..171K} and \citet{2017ApJ...837...91B} are also shown. }\label{Fig:energy} 
\end{figure}

To calculate the outflow size ($R_{\mathrm{out}}$), we used the empirical relation based on the integral field spectroscopy by \citet{2018arXiv180708356K}:
\begin{equation}
\log R_{\mathrm{out}} (\mathrm{kpc}) = (0.28\pm0.03) \times \log L_{\mathrm{[O\,III]}} - (11.27\pm 1.46).
\end{equation} 

The outflow velocity is difficult to measure due to various effects such as inclination, the geometry of the outflows, extinction and turbulent velocity, which need to be
carefully considered when the spatially integrated spectra as well as the spatially-resolved spectra are used for analysis. As a crude estimation, we combined velocity shift and dispersion as a representative velocity:
\begin{equation}
v_{\mathrm{out}}=\sigma_0 = \mathrm{\sqrt{{\sigma}^2_{[O\,III]} + V^2_{[O\,III]}}}
\label{eq:vout}
\end{equation} 

Using the estimated mass, size, and velocity, we determine the mass outflow rate ($\dot{M}_{\mathrm{out}}$), energy injection rates ($\dot{E}_{\mathrm{out}}$) and momentum flux ($\dot{P}_{\mathrm{out}}$) based on the following equations \citep[see][]{2017ApJ...837...91B}, which assume uniformly-filled spherical or biconical outflows \citep{2012MNRAS.425L..66M},
   
\begin{align}
\dot{M}_{\mathrm{out}}  = &  3 M_{\mathrm{gas}} \frac{v_{\mathrm{out}}}{R_{\mathrm{out}}} \\ 
\dot{E}_{\mathrm{out}} =  &  \frac{1}{2} \dot{M}_{\mathrm{out}} v^2_{\mathrm{out}}\\ 
\dot{P}_{\mathrm{out}} =  &  \dot{M}_{\mathrm{out}} v_{\mathrm{out}} 
\label{eq:Mout}
\end{align} 

In Figure \ref{Fig:energy}, we plotted $\dot{M}_{\mathrm{out}}$ (upper), $\dot{E}_{\mathrm{out}}$ (middle) and $\dot{P}_{\mathrm{out}}$ (lower) as a function of bolometric luminosity\footnote{Here bolometric luminosity was calculated from the [O III] luminosity as $L_{\mathrm{bol}} =3500 L_{\mathrm{[O\,III]}}$ following \citet{2004ApJ...613..109H} in order to make comparison with the literature samples.}. Energetics of gas outflows is found to strongly and positively correlate with bolometric luminosity. The Spearman's rank correlation coefficient ($r_s$) is found to be 0.62, 0.60 and 0.62, respectively with a $p$-value $<1\mathrm{e-20}$ in all cases. To compare our measurements with those reported in literatures based on the integral field spectroscopy or spatially resolved long-slit spectroscopy, we collected the outflow kinematics information of AGNs at $z<0.5$ from \citet{2011ApJ...732....9G}, \citet{2014MNRAS.441.3306H}, \citet{2016ApJ...833..171K} and \citet{2017ApJ...837...91B}. Note that these literature values are measured differently, e.g., using different emission lines, different equation for $v_{\mathrm{out}}$ etc. \citet{2016ApJ...833..171K} showed that H$\alpha$ based mass outflow rates are a factor of 10 larger than that of [O III] based. To have self-consistent measurements for all different samples, we calculated the outflow mass using total [O III] luminosity. Adopting outflow size, velocity dispersion, maximum velocity and electron density etc. from the respective papers, we calculated their $\dot{M}_{\mathrm{out}}$, $\dot{E}_{\mathrm{out}}$ and $\dot{P}_{\mathrm{out}}$ using the above equations. We caution that these estimated values for literature sources can be different from the values reported in the literature, mainly due to the use of [O III] luminosity in equation \ref{eq:mass} to calculate gas mass and the definition of outflow velocity used by different authors. For example, our estimated mass outflow rate and energy injection rate is a factor of 3 and 10 smaller, respectively, than the values estimated by \citet{2014MNRAS.441.3306H} using H$\beta$ \citep[see][]{2016ApJ...833..171K}. In addition, the difference in mass outflow rates is partly caused by differently assumed gas densities and the difference between the size of outflows compared to the extent of the ionized gas \citep{2018ApJ...856...46R}. The literature sources plotted in Figure \ref{Fig:energy} cover a wide range of AGN luminosities and outflow parameters. For example, \citet{2017ApJ...837...91B} have mostly low-luminosity AGNs, while \citet{2011ApJ...732....9G} include high-luminosity AGNs.

 The estimated $M_\mathrm{gas}$ for the Type 1 AGNs is found to have a range of $4.9 \times 10^3 - 2.0\times 10^{8} \, M_{\odot}$ with a median of $3.5\times 10^{5} \, M_{\odot}$. The outflow radius is ranging from 0.5 to 5.1 kpc with a median of $\sim 1.8$ kpc. At this radius, the estimated mass outflow rate is $0.01-126\, M_{\odot}\, \mathrm{yr^{-1}}$ with a median at $\sim 0.48 \, M_{\odot}\, \mathrm{yr^{-1}}$. The median mass accretion rate for the Type 1 AGNs calculated from $\dot{M}_{\mathrm{acc}}=L_{\mathrm{bol}}/\eta c^2$, assuming the accretion efficiency $\eta=0.1$ is $\sim 0.14 \, M_{\odot}\, \mathrm{yr^{-1}}$. Thus, the outflow rate for Type 1 AGNs is $\sim 1-100$ (with a median of $\sim 3$) times higher than the median mass accretion rate. Though majority of the Type 1 AGNs have mass loading factor (the ratio of outflow rate to the accretion rate) between $1-10$, powerful mass loading $>10$ is also seen in some AGNs (the upper dashed line) indicating powerful mass loading by the AGN outflow to the interstellar medium \citep[see][]{2005ARA&A..43..769V}. We note a positive correlation ($r_s \sim 0.3$) between Eddington ratio and mass outflow rate, i.e., high Eddington ratio sources have higher mass outflow rate. The mass outflow rate of our sample is consistent with literature samples over 3 orders of magnitudes in AGN luminosity. We note a large scatter in mass outflow rates at a given luminosity which is due to uncertainty in individual parameters that enter in the calculation such as outflow velocity (see Figure \ref{Fig:vvd_LOIII_Ledd}), outflow size and electron density.

The energy injection rates shown in the middle panel of Figure \ref{Fig:energy} for Type 1 AGNs are $2\times 10^{37}-8\times 10^{42}\, \mathrm{erg\, s^{-1}}$ with a median of $ 1 \times 10^{40}\, \mathrm{erg\, s^{-1}}$ consistent with the literature samples. Majority of our Type 1 sample are located below 0.01\% energy conversion efficiency line, indicating very low energy conversion efficiency and there is no source above 1\%. This is, in fact, similar to the literature samples, the majority of which are located below 0.01\%. The momentum flux range (lower panel of Figure \ref{Fig:energy}) is $5\times 10^{30}-2.8\times 10^{35}\, \mathrm{dyne}$ with a median of $8 \times 10^{32} \, \mathrm{dyne}$ for Type 1 AGNs. Thus, the momentum flux is also relatively low ranging between $0.01-0.1 \times L_{\mathrm{bol}}/c$, which is again consistent to the literature samples at similar luminosity range.

\section{Discussion}\label{sec:discussion}

The effect of kpc-scale outflows on the kinematics manifested by emission lines has been studied by various authors, in particular, high ionization [O III] line has been found to be a good tracer of outflows. We found that outflow indicators, i.e. $\sigma_{\mathrm{[O\,III]}}$ and V$_{\mathrm{[O\,III]}}$, strongly increase with AGN luminosity and Eddington ratio in both Type 1 and Type 2 AGNs, suggesting that ionized gas outflows are radiation driven. In $\sim 89$\% of the Type 1 AGNs, [O III] profile is well represented by double Gaussian profile indicating that the non-gravitational component is present in a majority of the Type 1 AGNs. This number is larger than that of the Type 2 AGNs which has $\sim 43$\% double Gaussian [O III] profile, mainly due to the average higher luminosity of Type 1 AGNs than Type 2 AGNs. We note that the fraction of AGNs with double Gaussian [O III] is sensitive to the quality of spectra and [O III] line strength because the wing in the [O III] profile may not be detected when the line is intrinsically weak or the quality of the spectrum is low. The previous studies reported a wide range of values from $\sim 25$\% to 70\% for Type 1 and Type 2 AGNs \citep[e.g.,][]{1995ApJS...99...67N,2008ApJ...680..926K,2017MNRAS.468..620Z}. The fraction of AGNs with double Gaussian [O III] steeply increases with AGN luminosity, suggesting the outflows are stronger in high-luminosity AGNs as also evident in the VVD diagram analysis. These results show that outflows are prevalent in both Type 1 and Type 2 AGNs, particularly in luminous AGNs.

Apart from the radiation/wind driven scenario, the interaction between AGN jets and the interstellar medium was suggested as a driver of gas outflows \citep[e.g.,][]{1984ApJ...281..525H,2004AJ....127..606W,2013MNRAS.433..622M,2014MNRAS.442..784Z}. \citet{2013MNRAS.433..622M} found a strong correlation between outflow velocity and radio luminosity concluding that the large-scale outflows are mainly driven by radio jet. A similar trend of increasing [O III] velocity dispersion with radio luminosity was also noticed in the radio-detected subsample of Type 2 AGNs by \citet{2016ApJ...817..108W}. However, the authors do not find any role of radio jets on the non-gravitational kinematics after normalizing [O III] velocity dispersion with stellar velocity dispersion. About 17\% of Type 1 AGNs of our sample have radio counterpart in FIRST radio survey, thus the majority of our AGNs do not have radio counterpart although they have non-gravitational component ($\sim 89$\% of our AGNs show double Gaussian [O III] profile). Among radio-detected AGNs, we see a positive correlation between velocity dispersion of [O III] and radio luminosity. However, such trend is weaker when plotted against radio loudness and become flat when total [O III] velocity dispersion is normalized to that of the narrow [O III] component. Our results agree with the findings by \citet{2018ApJ...852...26W}, who studied a sample of Type 1 AGNs having $z=0.4-0.8$. Thus, the non-gravitational component is not directly influenced by radio jet.

Recent advances in the integral field spectroscopic observation enabled us to study AGN energetics although limited to a handful number of local AGNs. Mass outflow, energy injection and momentum flux rates have been found to increase with AGN bolometric luminosity suggesting that the gas outflows are AGN driven \citep[e.g.,][]{2014MNRAS.441.3306H,2016ApJ...833..171K,2017ApJ...837...91B}. Theoretical models suggest photons originating close to the black hole initially drive an optically thick wind via radiation pressure \citep{2003MNRAS.345..657K}. However, large-scale outflows could be energy driven if outflowing gas expands adiabatically conserving its energy or momentum driven if gas loses energy after a short phase of radiative cooling \citep{2011MNRAS.415L...6K,2015ARA&A..53..115K}. Outflow within kpc-scale is believed to be momentum driven \citep{2016ApJ...833..171K} but become energy driven at large-scale \citep{2011MNRAS.415L...6K}.

We have quantitatively estimated mass outflow rates, energy injection rate and momentum flux for a large sample of Type 1 AGNs. We found a strong correlation of outflow energetics with AGN luminosity suggesting AGN driven outflow. The outflow size of AGNs in our sample is $0.5-5.1$ kpc with a median of 1.8 kpc. This implies the observed outflows in the majority of AGNs in our sample is momentum driven. According to \citet{2012MNRAS.425..605F} a fast wind ($>10,000 \, \mathrm{km \, s^{-1}}$) or even a slow wind of 1000 km s$^{-1}$ with some stringent conditions could lead to the energy conserving outflow. Since our Type 1 AGN sample lies between $0.01-0.1\, L_{\mathrm{bol}}/c$, energy-conserving outflows are unlikely. In the energy-conserving outflow \citep[see][]{2015Natur.519..436T} $\dot{P}_{\mathrm{out}} = f_c \times (v_{\mathrm{in}}/v_{\mathrm{out}}) \times (L_{\mathrm{bol}}/c)$, where $v_{\mathrm{in}}$ and $v_{\mathrm{out}}$ are the small-scale wind and large-scale outflow velocity, and $f_c$ is the fraction of the small-scale wind power that is transferred to the large-scale outflow. Assuming $f_c$ to be 0.2 \citep{2015Natur.519..436T}, most of the AGNs in our sample have $v_{\mathrm{in}}= (10^{-5} - 10^{-2})c$, which is much lower than the ultrafast outflow and unlikely to be driven by energy-conserving phase.

\section{Summary and conclusion}\label{sec:conclusion}
We have investigated ionized gas outflows based on the [O III] kinematics using a large sample of $\sim 5000$ Type 1 AGNs at $z<0.3$. 
For comparison, we combined Type 1 AGNs with the sample of $\sim 39,000$ Type 2 AGNs from \citet{2016ApJ...817..108W}. Our main findings are summarized as follows.
\begin{itemize}
\item For the majority of Type 1 AGNs ($\sim$89\%), the [O III] line profile presents a broad wing component, representing a non-virial motion, i.e., outflows. 
Compared to Type 2 AGNs, of which $\sim$43\% shows broad [O III] fitted with a double Gaussian model, outflow signature is more easily detected in Type 1 AGNs.
This is partially due to the luminosity effect since the mean luminosity of Type 1 AGNs is much higher than that of Type 2 AGNs. The fraction of AGNs with double Gaussian [O III] steeply increases with AGN luminosity and Eddington ratio in Type 1 AGNs as similarly found in Type 2 AGNs.

\item The velocity dispersion of [O III] strongly correlates with [O III] luminosity while Type 1 AGNs have on average higher velocity dispersion than Type 2 AGNs. 

\item Although many AGNs show $\sim$zero velocity shift, a significant fraction of AGNs presents strong velocity shift, suggesting various effects on the observed kinematic signatures, i.e., the inclination and opening angle of the cone and dust obscuration. The average velocity shift increases with the [O III] luminosity, while the velocity shift is larger in Type 1 AGNs than in Type 2 AGNs, reflecting the orientation and projection effect.  

\item The VVD diagram expands toward higher values with increasing AGN luminosity and Eddington ratio in both Type 1 and Type 2 AGNs, suggesting that outflows are radiation-driven.

\item Blueshifted [O III] is more frequently detected than redshifted [O III] in Type 1 AGNs as expected from the biconical outflow model combined with dust obscuration. 
The ratio between AGNs with blueshifted [O III] and AGNs with redshifted [O III] is higher in Type 1 AGNs than in Type 2 AGNs. These results are consistent with 
the unification model of AGNs and well reproduced by the 3D bicone models.

\item The apparent trend of increasing $\sigma_{\mathrm{[O\,III]}}$ with radio luminosity or radio-loudness vanishes when normalized by the velocity dispersion of [O III] core component, which is a proxy for the galaxy gravitational potential. This is in agreement with the previous findings for Type 2 AGNs where no trend of radio luminosity and dispersion have been found once [O III] velocity dispersion was normalized with stellar velocity dispersion, suggesting that outflows are not directly linked to the radio activity of AGNs.

\item Mass outflow, energy injection and momentum flux rates increase with AGN luminosity. A majority of the Type 1 AGNs in our sample have mass outflow rates between $1-10 \,L_{\mathrm{bol}}/\eta c^2$ indicating powerful mass loading by AGN outflows to the interstellar medium. Energy conversion efficiency is estimated to be smaller than 0.01\% of bolometric luminosity and momentum flux is ranging between $0.01-0.1  \,L_{\mathrm{bol}}/c$.

\end{itemize}

\acknowledgments 
We thank the anonymous referee for the suggestions, which improved the clarity of the manuscript. We are grateful to Hyun-Jin Bae for providing bicone model grid. SR thanks Jaejin Shin (SNU) and Neha Sharma (KHU) for useful discussion. This work has been supported by the Basic Science Research Program through the National Research Foundation of Korea government (2016R1A2B3011457). This work has made use of SDSS spectroscopic data. Funding for SDSS-III has been provided by the Alfred P. Sloan Foundation, the Participating Institutions, the National Science Foundation, and the U.S. Department of Energy Office of Science. The SDSS-III web site is http://www.sdss3.org/. SDSS-III is managed by the Astrophysical Research Consortium for the Participating Institutions of the SDSS-III Collaboration including the University of Arizona, the Brazilian Participation Group, Brookhaven National Laboratory, Carnegie Mellon University, University of Florida, the French Participation Group, the German Participation Group, Harvard University, the Instituto de Astrofisica de Canarias, the Michigan State/Notre Dame/JINA Participation Group, Johns Hopkins University, Lawrence Berkeley National Laboratory, Max Planck Institute for Astrophysics, Max Planck Institute for Extraterrestrial Physics, New Mexico State University, New York University, Ohio State University, Pennsylvania State University, University of Portsmouth, Princeton University, the Spanish Participation Group, University of Tokyo, University of Utah, Vanderbilt University, University of Virginia, University of Washington, and Yale University.

\software{mpfit \citep{2009ASPC..411..251M}, pPXF \citep{2004PASP..116..138C}}

 \bibliographystyle{apj}
 \bibliography{ref}

\end{document}